\documentclass[11pt,a4paper]{article}

\usepackage[T1]{fontenc}
\usepackage{eurosym}
\usepackage{booktabs}
\usepackage{multirow}
\usepackage{sectsty}
\usepackage{graphicx}
\usepackage{amsmath}
\usepackage{amssymb}
\usepackage{subcaption}
\usepackage{caption}
\usepackage{todonotes}
\usepackage{wrapfig}
\usepackage{hyperref}
\hypersetup{colorlinks=true,linkcolor=blue,urlcolor=blue}

\usepackage{fancyhdr}
\usepackage[left=2cm,top=1.5cm,bottom=1.5cm,right=2cm,includefoot,includehead,headheight=13.6pt]{geometry}
\usepackage{lastpage}
\usepackage{enumitem}
\usepackage{lipsum}
\usepackage[toc]{multitoc}

\numberwithin{equation}{section}

\newcommand{\simas}[1]{\raisebox{-.1ex}{
            $\stackrel{\small{#1}}{\sim}$}}

\newcommand{\ssfill}{\xleaders\hbox to 0.35em{\scriptsize.}\hfill}% d

\makeatletter
\newcommand{\oset}[3][0ex]{%
	\mathrel{\mathop{#3}\limits^{
			\vbox to#1{\kern-2\ex@
				\hbox{$\scriptstyle#2$}\vss}}}}
\makeatother

\newcommand*{\cventry}[7][.25em]{
  \noindent\begin{tabular*}{\textwidth}{l@{\extracolsep{\fill}}r}%
	  {\bfseries #4} & {\bfseries #5} \\%
	  {\itshape #3\ifthenelse{\equal{#6}{}}{}{, #6}} & {\itshape #2}\\%
  \end{tabular*}%
  \ifx&#7&%
    \else{\\\vbox{\small#7}}\fi%
  \par\addvspace{#1}}

\newcommand*{\hintfont}{\bfseries}
\newcommand*{\hintstyle}[1]{{\noindent\hintfont{#1}}}
\newcommand*{\cvitem}[3][.25em]{%
  \ifthenelse{\equal{#2}{}}{}{\hintstyle{#2}: }{#3}%
  \par\addvspace{#1}}

\let\OriginalQuotation\quotation
\renewcommand*{\quotation}{\OriginalQuotation\small\sf}

\begin{document}

\allsectionsfont{\sffamily}

% HEADERS
\fancyhead{}
\fancyfoot{}

\fancyhead[CO]{{}}
\fancyhead[LO]{{}}
\fancyhead[RO]{{}}
\fancyfoot[R]{\thepage\, / {\color[rgb]{0.6,0.,0}\pageref{LastPage}}}
\renewcommand{\headrulewidth}{0pt}

% Title
\newcommand*{\titleGP}{\begingroup % Create the command for including the title page in the document
\thispagestyle{empty}

\begin{flushright}
  DESY 20-127 \\
  IFIC/20-38
\end{flushright}

{\centering % Center all text

\begin{center}
{\LARGE {\sf An analysis of systematic effects in finite size scaling
    studies using the gradient flow }} \\[0.2\baselineskip]
\end{center}

%\rule{\textwidth}{0.4pt}\vspace*{-\baselineskip}\vspace{3.2pt} % Thin horizontal line
}
\begin{center}
  {\large Alessandro Nada$^a$ and Alberto~Ramos$^b$} 
\end{center}
\vspace{0.2cm}
\begin{center}
  $^a${John von Neumann Institute for Computing (NIC), DESY, Platanenallee~6, 15738~Zeuthen, Germany}\\
  $^b${Instituto de F\'{\i}sica Corpuscular (IFIC), CSIC-Universitat de Valencia, 46071, Valencia, Spain}
\end{center}

\vspace{1cm}
\begin{center}
  \large{\sf Abstract}
\end{center}
\rule{\textwidth}{0.4pt}
\noindent
We propose a new strategy for the determination of the step scaling
function $\sigma(u)$ in finite size scaling studies using the Gradient Flow. 
In this approach the determination of $\sigma(u)$ is broken in two
pieces: a change of the flow time at fixed physical size,
and a change of the size of the system at fixed flow time. 
Using both perturbative arguments and a set of simulations in the pure
gauge theory we show that this approach leads to a better control over
the continuum extrapolations. 
Following this new proposal we determine the running coupling at high
energies in the pure gauge theory and re-examine the determination of
the $\Lambda$-parameter, with special care on the perturbative
truncation uncertainties.
%%% Local Variables:
%%% mode: latex
%%% TeX-master: "paper"
%%% End:
\\\rule{\textwidth}{0.4pt}\\[\baselineskip] % Thin horizontal line

\tableofcontents

\newpage
\endgroup}

\begingroup % Create the command for including the title page in the document
\thispagestyle{empty}

\begin{flushright}
  DESY 20-127 \\
  IFIC/20-38
\end{flushright}

{\centering % Center all text

\begin{center}
{\LARGE {\sf An analysis of systematic effects in finite size scaling
    studies using the gradient flow }} \\[0.2\baselineskip]
\end{center}

%\rule{\textwidth}{0.4pt}\vspace*{-\baselineskip}\vspace{3.2pt} % Thin horizontal line
}
\begin{center}
  {\large Alessandro Nada$^a$ and Alberto~Ramos$^b$} 
\end{center}
\vspace{0.2cm}
\begin{center}
  $^a${John von Neumann Institute for Computing (NIC), DESY, Platanenallee~6, 15738~Zeuthen, Germany}\\
  $^b${Instituto de F\'{\i}sica Corpuscular (IFIC), CSIC-Universitat de Valencia, 46071, Valencia, Spain}
\end{center}

\vspace{1cm}
\begin{center}
  \large{\sf Abstract}
\end{center}
\rule{\textwidth}{0.4pt}
\noindent
We propose a new strategy for the determination of the step scaling
function $\sigma(u)$ in finite size scaling studies using the Gradient Flow. 
In this approach the determination of $\sigma(u)$ is broken in two
pieces: a change of the flow time at fixed physical size,
and a change of the size of the system at fixed flow time. 
Using both perturbative arguments and a set of simulations in the pure
gauge theory we show that this approach leads to a better control over
the continuum extrapolations. 
Following this new proposal we determine the running coupling at high
energies in the pure gauge theory and re-examine the determination of
the $\Lambda$-parameter, with special care on the perturbative
truncation uncertainties.
%%% Local Variables:
%%% mode: latex
%%% TeX-master: "paper"
%%% End:
\\\rule{\textwidth}{0.4pt}\\[\baselineskip] % Thin horizontal line

\tableofcontents

\newpage
\endgroup

\section{Introduction}

In recent years the gradient flow~\cite{Narayanan:2006rf,
  Luscher:2010iy} (GF) has found many successful applications. 
In particular, when combined with the ideas of finite-size
scaling~\cite{Luscher:1991wu}, it provides a powerful tool 
for the determination of the running coupling in asymptotically-free 
strongly-coupled gauge theories~\cite{Fodor:2012td, Fritzsch:2013je, Ramos:2014kla}.  
The applications include the determination of the strong
coupling in QCD, and the study of near conformal systems
(see~\cite{Witzel:2019jbe} for a recent review). 

Coupling definitions based on the gradient flow have some properties
that make them very attractive. 
First, the relevant observables show a small variance. 
A modest numerical effort allows their determination with a sub-percent
precision. Second, the gradient flow coupling is given directly as an
expectation value. Its determination only involves the numerical
integration of the flow equations, something that in practice can be
done with arbitrary precision without having to perform any fits, or
taking any limit.
This means that finite-size scaling studies using the GF only have one
source of systematic effect: the continuum extrapolation.

Nevertheless, it has been shown that these systematic effects are 
difficult to keep under control (see~\cite{Ramos:2015dla} for a review). 
It was soon noticed~\cite{Lin:2014fxa} that seemingly innocent
extrapolations could cause large systematic effects. 
Although the same work suggested to simply use larger flow times as a
way to have a better control of the continuum extrapolation, this
comes with an increase in the variance of the observable. 
One of the strong points of GF studies, the high statistical
precision, had to be sacrificed. 
Some efforts were made in order to understand the anatomy of these cutoff
effects, first in a perturbative context to tree
level~\cite{Fodor:2014cpa}, and later more systematically from an
effective field theory point of view~\cite{Ramos:2015baa}. 
It became clear that the integration of the flow equations and the
evaluation of the relevant observables at positive flow time could be
performed in a way that no $\mathcal O(a^2)$ effects are produced:
it is enough to use \emph{classically improved}
discretizations~\cite{Ramos:2015baa}. 
Still, the use of these
improved observables did not reduce substantially the observed scaling
violations (see for example~\cite{DallaBrida:2016kgh}). 
% An additional counterterm, related with the initial condition of the
% flow equation, and that only affects the scaling violations of flow
% observables is most probably the responsible for this large cutoff
% effects (see~\cite{Husung:2019ytz}). 
% Even if this picture is finally
% confirmed, there is no known way to determine this coefficient
% non-perturbatively. 
Unavoidably large lattices have to be simulated in
GF studies, and the continuum extrapolation will remain the main source of concern. 

In this work we study the main sources of systematic effects in finite-size 
scaling studies using the GF. 
We will argue that \emph{changes in the flow time} $t$ are
responsible for the scaling violations. 
On the contrary, when different finite volume couplings are determined \emph{at
the same value of the flow time} $t$, the scaling violations are very small. 
This observation will be supported by a non-perturbative study. 
Moreover it will allow us to propose a new strategy for the
determination of the step scaling function by breaking it up
into two pieces: first a change in the flow time,
without any change in the volume, second a change in the volume
without any change in the flow time. 
Since the first step can be performed \emph{without having to double
  the lattices}, the continuum extrapolation can be performed much
more accurately. We will discuss in detail the advantages of this approach.
Finally we will apply this alternative strategy to the case of pure gauge ${\rm SU}(3)$. 
Using the same datasets of~\cite{DallaBrida:2019wur}, we will revisit
the most crucial part of this work: the running at high energies and
the matching with the asymptotic perturbative regime. 

We also include an analysis of the variance of flow observables, 
allowing us to predict the dependence of the statistical uncertainties
with the flow time $t$.

%%% Local Variables:
%%% mode: latex
%%% TeX-master: "paper"
%%% End:

\section{The continuum extrapolation}

\subsection{Preliminaries}

\begin{figure}
  \centering
  \includegraphics[width=0.5\textwidth]{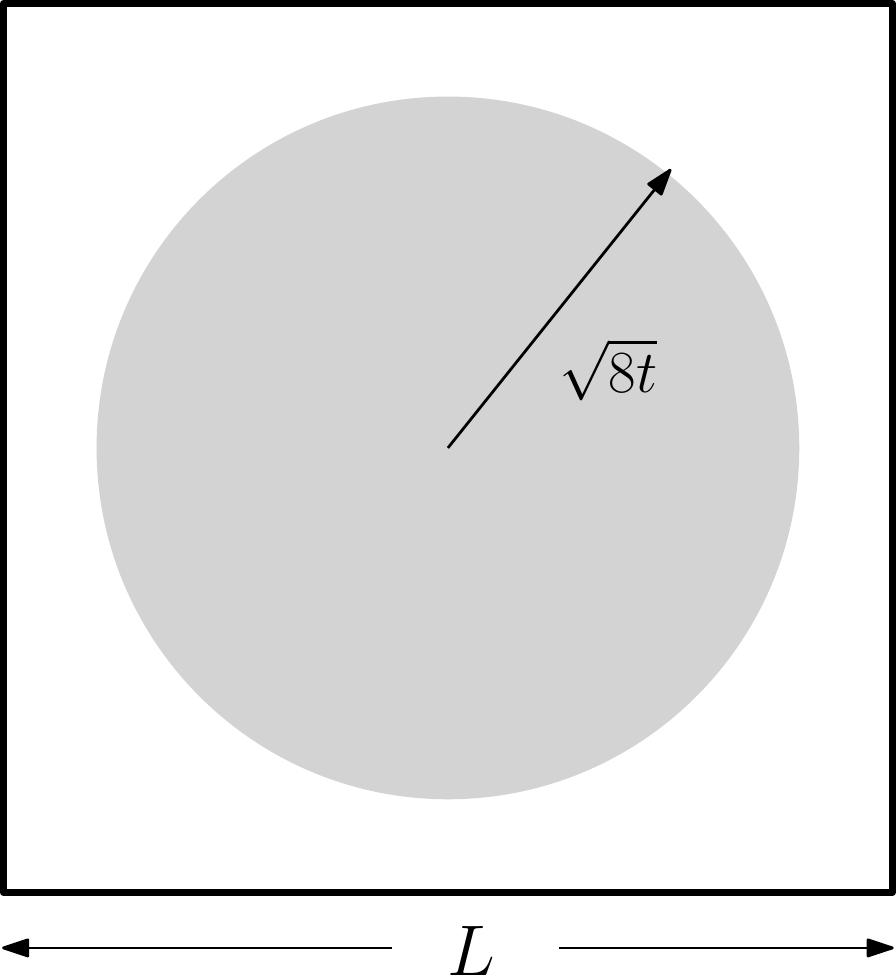} 
  \caption{The GF coupling in finite volume $\bar g ^2(\mu)$ is measured by computing
    the action density of the flow field $B_\mu(t,x)$ smeared over a
    distance $\sqrt{8t} = \mu^{-1}$ (see eq.~(\ref{eq:gsq})). 
  The renormalization scale $\mu$ and the size of the system $L$ are
  linked by the relation $\sqrt{8t} = cL$ (with $c$ a constant).}
  \label{fig:gsq}
\end{figure}

The gradient flow provides a set of renormalized observables with
small variance that are easy to compute via numerical simulations. 
The key idea consists in adding an extra time coordinate to the
gauge field, the flow time $t$. %, with dimensions of length squared. 
The dependence of the gauge field $B_{\mu}(t,x)$ on the flow time is
given by the first order diffusion equation 
\begin{eqnarray}
\label{eq:flow_cont}
  \partial_t B_{\mu}(t,x) = D_{\nu}G_{\nu\mu}(t,x)\,;&& \qquad B_{\mu}(0,x) = A_\mu(x)\,. \\
  G_{\mu\nu}(t,x) = D_{\mu}B_{\nu}(t,x) - D_{\nu}B_{\mu}(t,x)\,;&&\qquad D_\mu = \partial_\mu + [B_\mu,\cdot]\,.
\end{eqnarray}
The flow time has dimensions of length squared, and therefore
introduces a \emph{scale} into the problem (see figure~\ref{fig:gsq}). 
Gauge-invariant composite operators
defined at positive flow time are automatically renormalized~\cite{Luscher:2011bx}. 
In particular, a renormalized coupling at scale $\mu = 1/\sqrt{8t}$
can be defined by using the action density~\cite{Luscher:2010iy}
\begin{equation}
  \label{eq:gsq}
  \bar g ^2(\mu)\Big|_{\mu = 1/\sqrt{8t}} \propto t^2 \langle E(t,x) \rangle\,.\qquad \left( E(t,x) = G_{\mu\nu}^a(t,x)G_{\mu\nu}^a(t,x)  \right)
\end{equation}
In finite volume schemes where the invariance under Euclidean time 
translations is broken, like
the Schr\"odinger Functional (SF) or open-SF boundary conditions, the
coupling is only measured at the time-slice $x_0=T/2$ and usually
only certain components of the field strength are used 
\begin{subequations}
  \label{eq:Ecomp}
  \begin{eqnarray}
  E_{\rm m}(t,x) &=& G_{ij}^a(t,x)G_{ij}^a(t,x)\Big|_{x_0=T/2} \,,\\
  E_{\rm e}(t,x) &=& G_{0j}^a(t,x)G_{0j}^a(t,x)\Big|_{x_0=T/2} \,.
  \end{eqnarray}
\end{subequations}
The coupling defined using $E=E_{\rm m}$ is usually referred as
\emph{magnetic} and the one using $E=E_{\rm e}$ as \emph{electric}. 

Most applications of the gradient flow until today derive from these
coupling definitions. 
In infinite volume the scale $\mu_0 = 1/\sqrt{8t_0}$ at which the
coupling $\bar g ^2(\mu_0)$ takes some pre-defined value is used to
define the reference scale $t_0$. 
In the context of finite-size scaling, the renormalization scale $\mu
= 1/\sqrt{8t}$ is linked with the linear size of the system
\begin{equation}
  \label{eq:cdef}
  \mu^{-1} = \sqrt{8t} = cL\,,
\end{equation}
and therefore the coupling ``runs'' with the size $L$ (see fig~\ref{fig:gsq}). 
The constant $c$ defines a scheme by fixing the ratio of the length of
the system and the flow time scale $\sqrt{8t}$. 
The full definition of the GF coupling in finite volume
reads
\begin{equation}
  \label{eq:gsqfv}
  \bar g ^2_c(\mu) = \mathcal N(c)\,  t^2 \langle E(t,x) \rangle\Big|_{\mu^{-1} = \sqrt{8t} = cL}\,.
\end{equation}
The constant $\mathcal N(c)$ has been determined for
different choices of boundary conditions~\cite{Fodor:2012td,
  Fritzsch:2013je, Ramos:2014kla, Luscher:2014kea}. 

When computing the gradient flow coupling on the lattice there are
several choices to make, beyond the action chosen for the simulation. 
First, the flow equation~(\ref{eq:flow_cont}) has to be
translated to the lattice. Second, there are many valid
discretizations of the energy density $E(t,x)$. 
It is well understood that using the \emph{Zeuthen} flow to integrate
the flow equations and using any classically improved discretization
of the action density $E(t,x)$ guarantees that no further
$\mathcal O(a^2)$ effects are generated. 
The only remaining $\mathcal O(a^2)$ effects are those of the choice
of action for the simulation and an additional counterterm, only
affecting flow quantities~\cite{Ramos:2015baa}. 

On the lattice, where we can work only with dimensionless quantities,
the finite volume gradient flow coupling (equation~(\ref{eq:gsqfv}))
can be measured by determining $\bar g ^2(\mu)$ at the flow time (in
lattice units) $\sqrt{8t} / a = cL/a$. It is also common to use a
lattice version of the normalization factor $\mathcal N(c,a/\sqrt{8t}
)$ instead of the $\mathcal N(c)$ of eq.~(\ref{eq:gsqfv}),
in order to ensure that the leading order perturbative relation  
\begin{equation}
  \bar g ^2_c(\mu) \simas{\mu\to\infty} \bar g ^2_{\overline{\rm MS}}(\mu) + \mathcal O(\bar g ^4_{\overline{\rm MS}})\,,
\end{equation}
is exact for any lattice size.

The key character in finite-size scaling studies is the step scaling
function
\begin{equation}
  \sigma(u) = \bar g_c ^2(\mu/2) \Big|_{\bar g_c ^2(\mu) = u}\,.
\end{equation}
It measures how much the coupling changes under a variation in the
renormalization scale of a factor two\footnote{Other scale factors,
like 3/2, are less common, but obviously possible. 
The discussion in this paper also applies to any other choice.}. 
Its determination in lattice simulations is performed via a matching
of the bare parameters. 
At a fixed value of the bare coupling $g_0^2$ (and therefore fixed
value of the lattice spacing $a$)\footnote{One also needs to specify
the value of the quark masses. Simulations on a finite volume make it
possible to directly simulate at $m=0$, and this is typically the 
additional condition that completes the line of constant physics.}, 
one determines the GF coupling on a lattice of size $L/a$ (resulting 
in $\bar g ^2(\mu)$), and on a lattice $2L/a$ (resulting in $\bar g ^2(\mu/2)$). 
This allows the determination of a lattice approximation of the step scaling
function:
\begin{equation}
\label{eq:lattice_sigma}
  \Sigma(u, a/\sqrt{8t}) = \bar g_c ^2(\mu/2) \Big|_{\bar g_c ^2(\mu) = u}
  \simas{a/\sqrt{8t} \to 0} \sigma(u)\,.
\end{equation}
In the rest of this section we will be concerned with the continuum
extrapolation of $\Sigma(u,a/\sqrt{8t})$; before that however, let us insist on
two points:
\begin{itemize}
\item The direct determination of $\Sigma$, as suggested above, requires the
determination of the GF couplings \emph{at two different renormalization
  scales} $\mu$ and $\mu/2$, on lattices \emph{of two different sizes}
$L/a$ and $2L/a$. 

\item It is clear that once $c$ (eq.~(\ref{eq:cdef})) is fixed, taking
  $a/L\to 0$ is equivalent to taking $a/\sqrt{8t} \to 0$, so the
  usual notation that uses $a/L$ as the variable to parametrize cutoff
  effects is fully justified. 
  Nevertheless, it is $a/\sqrt{8t} = a\mu$ the natural variable to
  measures the size of cutoff effects. 
  Note that for the typical choices of $c\in [0.2-0.5]$ we have scale
  $\sqrt{8t} < L$.
  
  With this choice of the renormalization scale, it is clear that, at fixed $L/a$,
  larger values of $c$ would lead to \emph{smaller} cutoff
  effects. This notation will also be convenient for the discussion
  that follows.
\end{itemize}

\subsection{A new strategy for the determination of $\sigma(u)$}

\label{sec:new-strat-determ}

\begin{figure}
  \centering
  \includegraphics[width=0.5\textwidth]{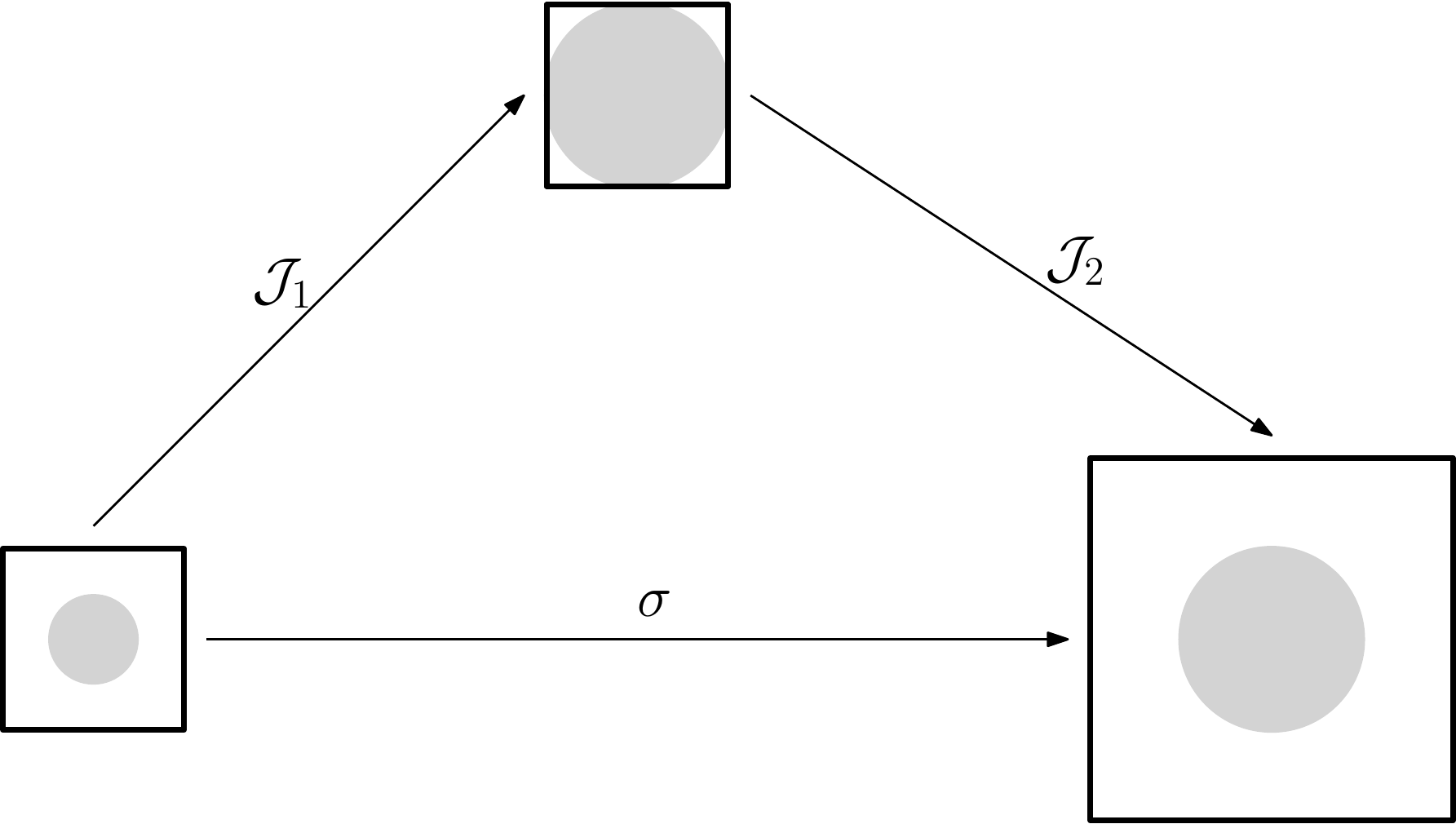} 
  \caption{The determination of the step scaling function $\sigma$
    involves a change in the renormalization scale and in the size of the
    system of a factor two.
    These two steps do not need to be performed at the sime time. 
    The figure shows that $\sigma$ can be determined as the
    composition of the function $\mathcal J_1$ that changes the
    renormalization scale $\mu \to \mu/2$ at fixed size $L$, 
    with the function $\mathcal J_2$ that changes the size 
    $L\to 2L$ at fixed renormalization scale $\mu$.}
  \label{fig:sigma}
\end{figure}

As already pointed out, the determination of the lattice step scaling
function involves two steps: a change in the renormalization scale by
a factor two, and change in the lattice size by the same factor. 
In previous works these two changes have been performed at the same time 
in a single step, but conceptually there is no need to do so.

Figure~\ref{fig:sigma} shows that the value of the step scaling
function $\sigma(u)$ can be determined by the composition of two functions. 
First we have
\begin{equation}
\label{eq:J1def}
  \mathcal J_1 (u) = \bar g ^2_{2c}(\mu/2)\Big|_{\bar g ^2_c(\mu) = u}\,.
\end{equation}
This function changes the renormalization scale by a factor two $\mu\to\mu/2$ at
fixed physical size $L$. 
Second, we need to determine
\begin{equation}
\label{eq:J2def}
  \mathcal J_2 (u) = \bar g ^2_{c}(\mu/2)\Big|_{\bar g ^2_{2c}(\mu/2) = u}\,. 
\end{equation}
This function changes the lattice size $L \to 2L$ keeping constant the
renormalization scale $\mu^{-1} = \sqrt{8t}$.

The relation
\begin{equation}
  \label{eq:sigma_comp}
  \sigma = \mathcal J_2 \circ \mathcal J_1\,,
\end{equation}
is now exact, and provides an alternative method to determine $\sigma$.

Our main assumption is that \emph{large scaling violations 
come with changes in the renormalization scale}. 
We will later provide evidence that this is the case, but for
the moment let us discuss why this opens up the possibility to improve 
the quality of the continuum extrapolations.
Let us start by explaining how these functions are computed in practice:
\begin{itemize}
\item The determination of $\mathcal J_1$ involves measuring how much
  the coupling changes when the renormalization scale is varied as $\mu
  \to \mu/2$ at constant physical size $L$. 
  This is simply achieved by measuring on a lattice simulation the
  value of the GF coupling at two different flow times (i.e. 
  $t\to 4t$, see eq.~(\ref{eq:cdef})). 
  Crucially, this determination does not require to double the lattice
  size, allowing precise results without the need of very large values of $L/a$.
  
\item The determination of $\mathcal J_2$ requires to change the
  physical size $L$ without varying the renormalization scale. 
  In practice one fixes the bare coupling $g_0^2$ at a given value, and
  then measures the GF coupling on a $L/a$ lattice at flow
  time $\sqrt{8t}/a = cL/a$ and on a $2L/a$ lattice at the same $t$.
  This step requires to change the lattice size, but since the
  renormalization scale remains the same, one expects reduced cutoff effects. 
  In some sense the determination of $\mathcal J_2$ corresponds to
  measuring the finite volume effects in $\bar g ^2_c(\mu)$ (see figure~\ref{fig:sigma}).
\end{itemize}

% In summary, such a breaking up of the computation might be convenient, because the
% determination of $\mathcal J_1$ involves no doubling of the lattice
% size, and large ranges of lattice spacing $a$ can be explored in the
% continuum extrapolation. 
% The doubling of the lattice size is performed when determining
% $\mathcal J_2$, but in this case we expect very small scaling
% violations, since the renormalization scale is not changed. 
In the rest of this section we will provide evidence of these statements,
but before that let us further comment on two points:
\begin{itemize}
\item The determination of the functions $\mathcal J_1$ and 
  $\mathcal J_2$ can be done on the lattice just by applying the
  definitions equation~(\ref{eq:J1def}) and~(\ref{eq:J2def}). 
  Note however that in this case the functions will carry a dependence
  on the cutoff. 
  We will label these functions $\hat{\mathcal J}_1(u,a/\sqrt{8t})$
  and $\hat{\mathcal J}_2(u,a/\sqrt{8t})$ respectively.
\item In this work all numerical results make use of the same discretizations
used in~\cite{DallaBrida:2019wur}. 
We encourage the reader to consult this reference for more details. 
Here it is enough to say that we define the GF
coupling with SF boundary conditions and that our preferred setup, based on
theoretical expectations, uses the Zeuthen flow and an improved
definition of $E(t,x)$. This preferred setup will also be compared with
the more common combination Wilson flow/Clover observable. 

Moreover, we will focus in the rest of the text on the \emph{magnetic} 
definition of the GF coupling (see
equations~(\ref{eq:Ecomp}))\footnote{Reference~\cite{DallaBrida:2019wur}
showed a perfect agreement between the electric and magnetic schemes. 
Focusing on one choice keeps the discussion and the notation more
simple.}. Of course, our discussion is general, and does not depend on
these particular choices. 
\end{itemize}

\subsubsection{Leading order perturbation theory}
\label{sec:perturbation-theory}

As a first look at the proposal we examine the leading
order perturbative relation. We use the continuum
norm $\mathcal N(c)$ in the evaluation of the finite volume couplings
$\bar g ^2_c$ and examine, to leading order in perturbation theory,
the quantities 
\begin{equation}
\hat{\mathcal C_0}(a/\sqrt{8t}) = \frac{\Sigma(u,a/\sqrt{8t} )}{u}\,, \quad
\hat{\mathcal C}_1(a/\sqrt{8t}) = \frac{\hat{\mathcal J}_1(u,a/\sqrt{8t})}{u}\,, \quad
\hat{\mathcal C}_2(a/\sqrt{8t}) = \frac{\hat{\mathcal J}_2(u,a/\sqrt{8t})}{u}\,.
\end{equation}
Note that since we are working at leading order in $\bar g ^2$, and
thanks to the normalization by the constant factor $u$, all
these quantities are one in the continuum.  

In this example we will examine a typical case where we consider
data for
\begin{equation}
  L/a=8, 10, 12, 16, 18, 20, 24, 32, 36, 40, 48\,.
\end{equation}
We will use $c=0.2$ (see eq.~(\ref{eq:cdef})).
Let us note a few basic points:
\begin{itemize}
\item The determination of $\hat{\mathcal C_0}(a/\sqrt{8t})$ and $\hat{\mathcal
  C_2}(a/\sqrt{8t} )$ requires to double the lattice size. 
  This means that with our data, lattice estimates for these
  functions will be available only for a factor 3 change from the finest to the coarsest 
  lattice spacings: $8\to 16, 10\to 20, 12\to 24, 16\to 32, 18\to 36, 20\to 40,
  24\to 48$.
\item On the other hand, the determination of $\hat{\mathcal
    C_1}(a/\sqrt{8t})$ only requires the measurement of the GF coupling at
  different values of the flow time $t$. 
  This can be done on all lattices, and our dataset provides a factor
  $6$ change from the coarsest ($L/a=8$) to the finest ($L/a=48$) lattice spacing.
\end{itemize}

\begin{figure}
  \centering
  \includegraphics[width=0.8\textwidth]{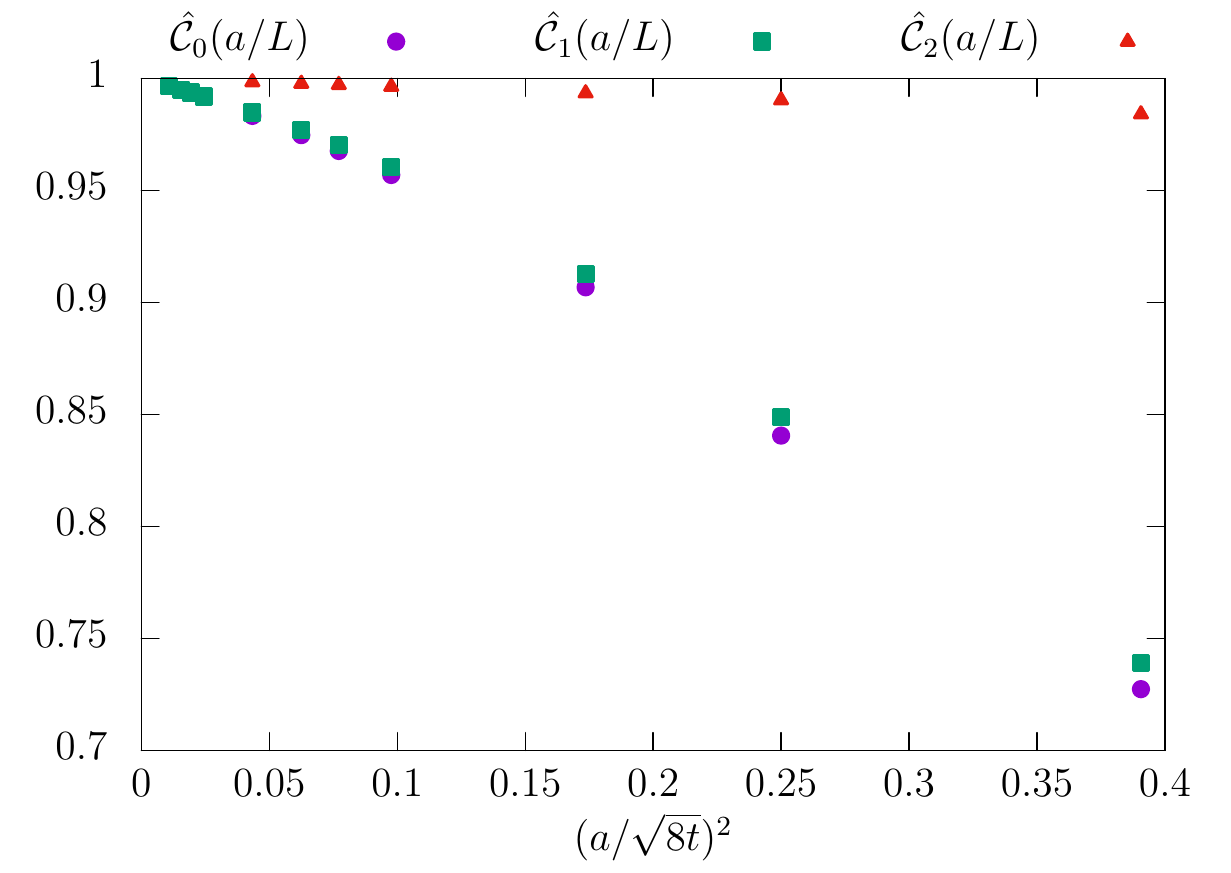} 
  \caption{Cutoff effects in the usual step scaling function
    ($\hat{\mathcal C_0}(a/\sqrt{8t} )$), compared to those in
    $\mathcal J_1$ (see $\hat{\mathcal C_1}(a/\sqrt{8t} )$) and
    $\mathcal J_2$ (see $\hat{\mathcal C_2}(a/\sqrt{8t} )$). See text for
    more details. Here we show the case of the Zeuthen flow/improved
    observable with plaquette gauge action (i.e. 
    the same setup that will be used in our non-perturbative study).}
  \label{fig:pt}
\end{figure}

Figure~\ref{fig:pt} shows the perturbative results. 
As the reader can see, the cutoff effects in $\hat{\mathcal
  C_0}(a/\sqrt{8t} )$ are very similar to those in $\hat{\mathcal C_1}(a/\sqrt{8t} )$. 
This can be understood in a simple way, since both these functions 
involve a change in the renormalization scale by a factor two. 
The main difference between both cases is that $\hat{\mathcal
  C_1}(a/\sqrt{8t} )$ can be determined using lattice spacings 
that are a factor two smaller, since
its determination does not require any change in the lattice size. 
The determination of $\hat{\mathcal C_2}(a/\sqrt{8t} )$ does not
involve any change in the renormalization scale, and it shows cutoff
effects that are one order of magnitude \emph{smaller} than either
$\hat{\mathcal C_0}(a/\sqrt{8t} )$ or $\hat{\mathcal C_1}(a/\sqrt{8t} )$. 

In the next section we will show that indeed these properties hold
non-perturbatively, and that they are not a coincidence of leading order
perturbation theory. 

\subsubsection{Non-perturbative study}
\label{sec:non-pert-study}

In this section we will describe the non-perturbative results used in
this study, first to support the claim that the numerical determination of 
$\hat{\mathcal J}_2$ has very small scaling violations, due to the fact that
the renormalization scale is not changed (i.e. 
the determination of $\hat{\mathcal J}_2$ amounts to measuring finite volume
effects in the coupling). Then, we want to show that
the determination of $\mathcal J_1$ can be performed accurately even
at values of $c$ that are too small to allow for a conservative estimate
of the step scaling function. 

All the analysis have been performed using two different analysis codes: 
one~\cite{Ramos:2018vgu} based on the
$\Gamma$-method~\cite{Ramos:2018vgu, Wolff:2003sm, Madras1988, Virotta2012Critical}, 
and the other using a jackknife resampling technique. 
Both analysis techniques take into account the correlations between
observables measured on the same ensemble. This is crucial, both
for the determination of $J_1$ that involves the measurements of $\bar g^2$ on
the same configuration at different values of the flow time, and also
to correctly determine the uncertainty in the composition $J_2 \circ J_1$.

\bigskip
\noindent
{\em Description of the data set}
\medskip

For our non-perturbative study we are going to use exactly the same dataset 
of~\cite{DallaBrida:2019wur}. This setup includes simulations of
the pure gauge theory with the Wilson plaquette action on
a lattice of size $L^4$ and lattice spacing $a$. 
We have several resolutions, $L/a=8, 10, 12, 16, 20, 24, 32, 48$, at a
large range of lattice spacings $a$ and with Schr\"odinger Functional (SF) boundary
conditions. The setup is the same as the one used for the perturbative study
in section~\ref{sec:perturbation-theory}.

We have measurements of the GF coupling at values of $c=0.2, 0.3,
0.4$ with two different discretizations: the usual Wilson flow/Clover
observable and the Zeuthen flow/improved observable (more details can be found
in~\cite{DallaBrida:2019wur}). Our target will be to determine the
running non-perturbatively in the scheme defined by $c=0.2$ by
computing the associated step scaling function $\sigma(u)$. 
Note that in reference~\cite{DallaBrida:2019wur} the value $c=0.3$ was
used because the large scaling violations at $c=0.2$ did not allow for
a determination of $\sigma(u)$.

We will revisit this attempt at a direct determination of $\sigma(u)$ here. 
Moreover, together with the data at $c=0.4$, we will be able to
determine both $\mathcal J_1, \mathcal J_2$, and compare their
composition with the direct determination of $\sigma$. 

Finally, the data with $c=0.3$ will be used in
section~\ref{sec:high-energy-regime} to compare the results
of~\cite{DallaBrida:2019wur} (where the $\Lambda$ parameter is
obtained by using a direct determination of the step scaling function
with $c=0.3$) with our new strategy. 

We will focus our investigations on the high energy regime, where
$\bar g ^2\sim 1-3$. 
Note that this region showed significant scaling violations, and in
fact turns out to be the most 
delicate part of the analysis in the extraction of $\Lambda$
(see~\cite{DallaBrida:2019wur} for more details).

\bigskip
\noindent
{\em Scaling violations in $\mathcal J_2$}
\medskip

\begin{figure}
  \centering
  
\begin{subfigure}[t]{0.49\textwidth}
  \includegraphics[width=\textwidth]{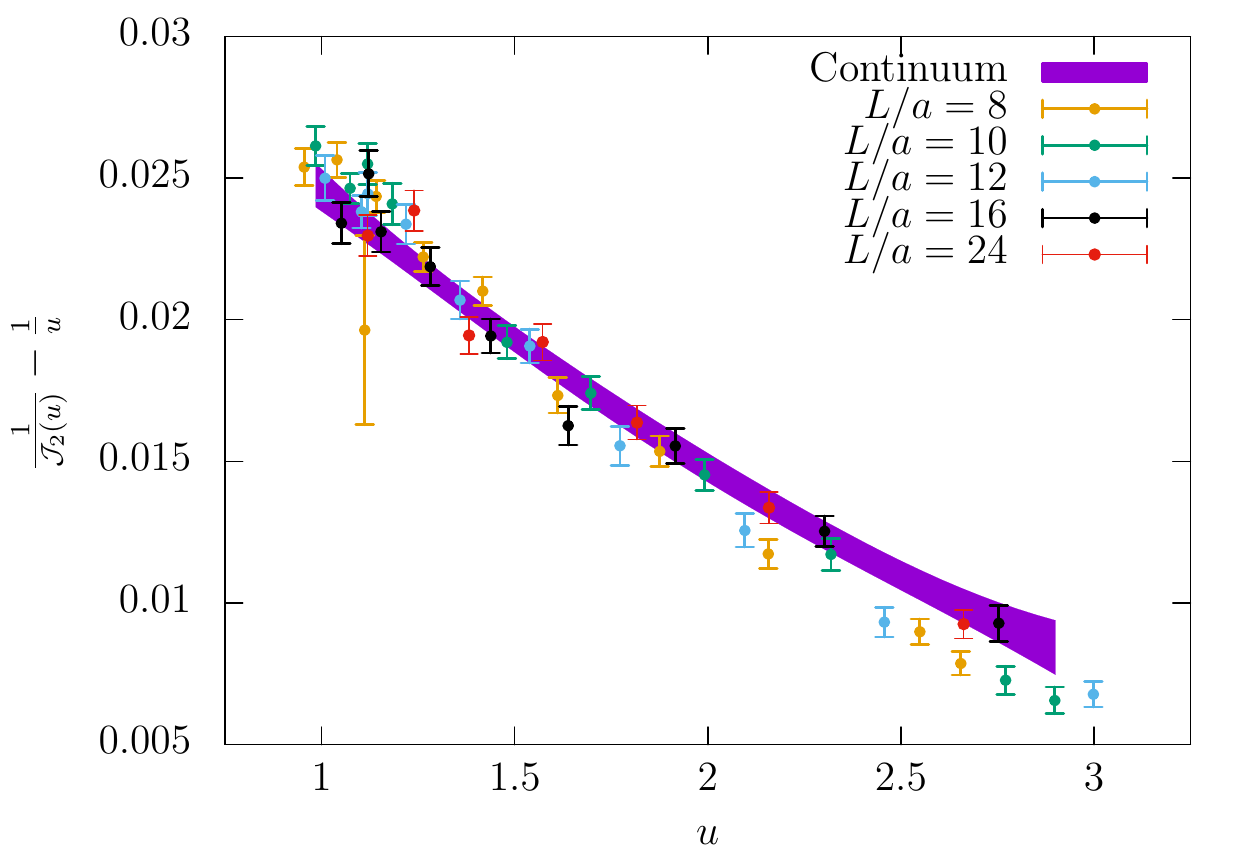} 
  \caption{Results for the combination of eq.~(\ref{eq:comb_J2}) and continuum extrapolation from global fit.}
  \label{fig:plot_J2}
\end{subfigure}
\begin{subfigure}[t]{0.49\textwidth}
  \includegraphics[width=\textwidth]{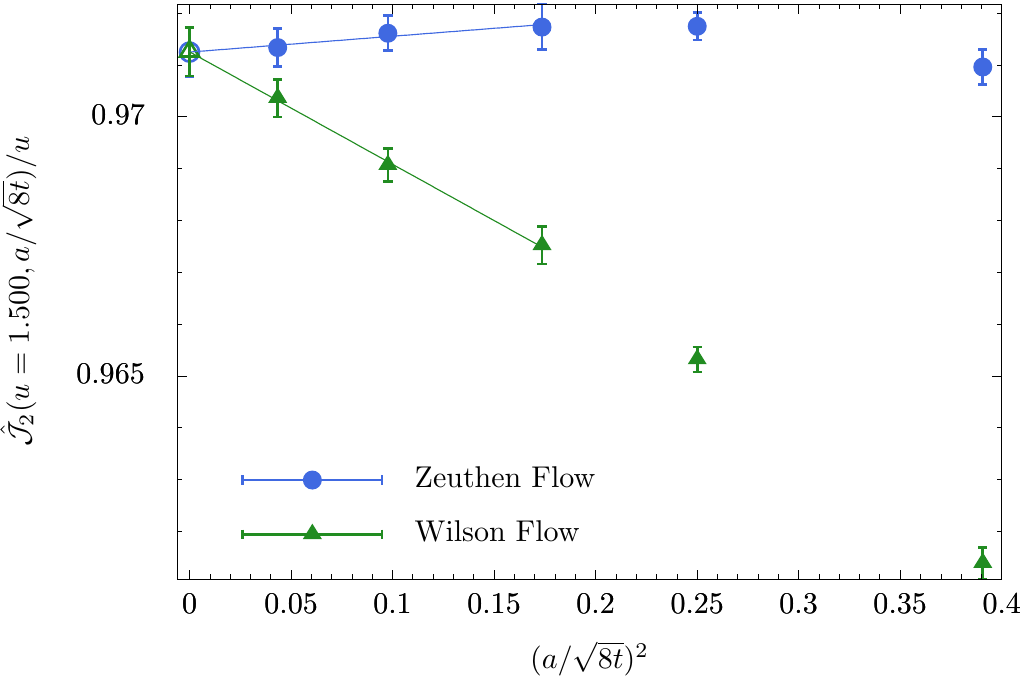}
  \caption{Results for the $\hat{\mathcal J_2} (u,a/\sqrt{8t})/u$ ratio for $u=1.5$ for different discretizations. The continuum extrapolation is also included.}
  \label{fig:J2_uv2}
\end{subfigure}
\caption{Results for the $\hat{\mathcal J_2} (u,a/\sqrt{8t})$ function and continuum extrapolation.}
\label{fig:J2o2}
\end{figure}

Let us start by investigating the scaling violations of $\mathcal J_2$. 
It is convenient to study the combination
\begin{equation}
\label{eq:comb_J2}
  \frac{1}{\hat{\mathcal J}_2(u, a/\sqrt{8t})} - \frac{1}{u} =  f_2(u, a/\sqrt{8t} )\,.
\end{equation}
The continuum limit of the right hand side, $f_2(u,0)$, has an
asymptotic expansion (in perturbation theory) as a polynomial
in $u$, starting with a constant term. 
Note that our data set allows to determine
$\hat{\mathcal J}_2(u,a/\sqrt{8t})$ for a factor three change in $a/\sqrt{8t}$.

The numerical raw data for $\hat{\mathcal J}_2(u,a/\sqrt{8t})$ is shown in 
figure~\ref{fig:plot_J2}. We also include in the plot the continuum
extrapolation.  
At this point we defer the discussion on how this continuum
curve is determined to section~\ref{sec:high-energy-regime}, and focus
on the key element: the non-perturbative data for $\hat{\mathcal J}_2$ show
very small scaling violations for all values of the coupling under study. 
The continuum curve is at most two standard deviations away from the
coarser lattice data ($\sqrt{8t}/a \approx 1.6$), and the two finest
lattices with $\sqrt{8t}/a \approx 3.2, 4.8$ show no significant
deviation from the continuum value.  

One can look in more detail at the previous statement by interpolating
the data with different $L/a$ to a common value of $u$, and then look
at the continuum extrapolation of $\hat{\mathcal J}_2$. 
We choose the value $u=1.5$, where we have several points at each
$L/a$, and therefore the necessary interpolations can be performed 
in the safest conditions available. %without incurring in any significant systematic. 
Figure~\ref{fig:J2_uv2} shows that the Zeuthen flow data shows no
significant scaling violations in the whole range of lattice spacing. 
The Wilson flow data show some scaling violations, but they are rather
mild, with the finest lattice being almost compatible with the continuum value. 
Extrapolations of the data with both discretizations are in full
agreement with each other.
%Extrapolations of the data with both discretizations are in full
%agreement with the result of the continuum curve of
%figure~\ref{fig:plot_J2}, labeled \emph{Global} (see
%section~\ref{sec:high-energy-regime} for more details on the
%determination of these values).

\bigskip
\noindent
{\em Scaling violations in $\mathcal J_1$}
\medskip

\begin{figure}
  \centering
  
\begin{subfigure}[t]{0.49\textwidth}
  \includegraphics[width=\textwidth]{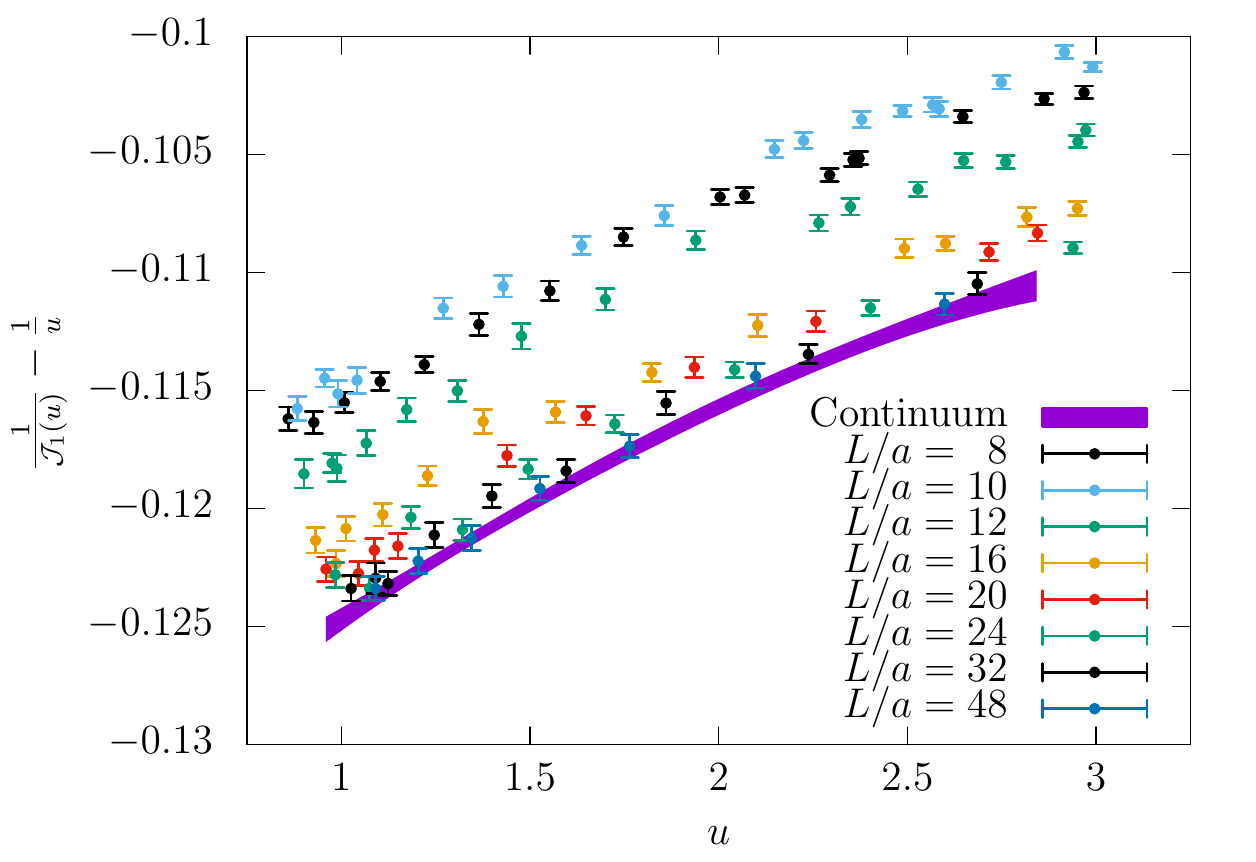} 
  \caption{Results for the combination of eq.~(\ref{eq:comb_J1}) and continuum extrapolation from global fit.}
  \label{fig:plot_J1}
\end{subfigure}
\begin{subfigure}[t]{0.49\textwidth}
  \includegraphics[width=\textwidth]{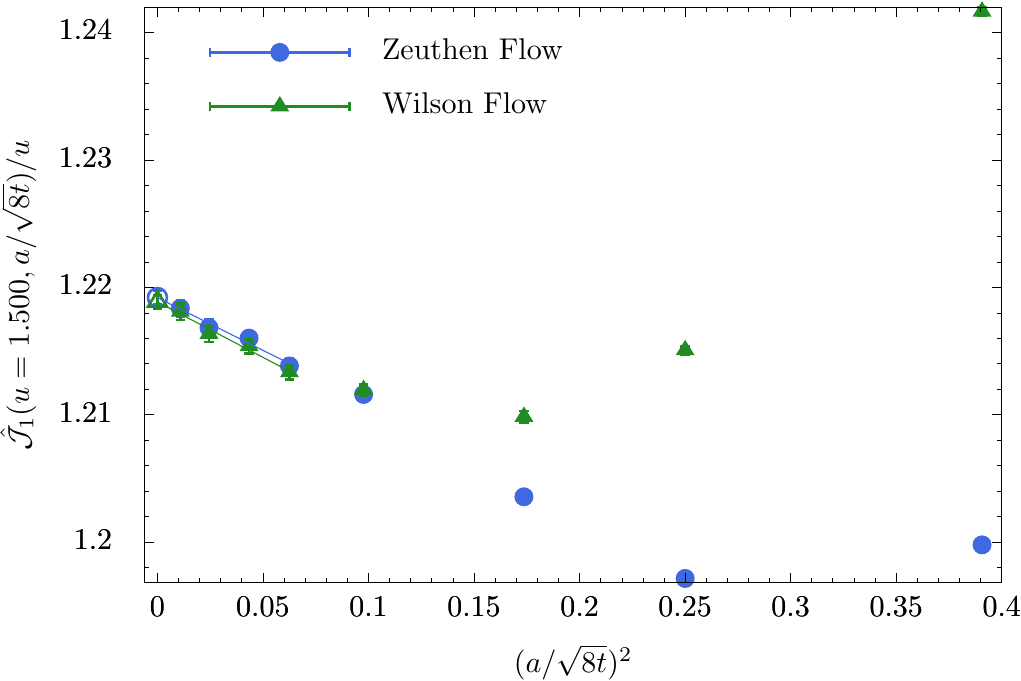}
  \caption{Results for the $\hat{\mathcal J_1} (u,a/\sqrt{8t})/u$ ratio for $u=1.5$ for different discretizations. The continuum extrapolation is also included.}
  \label{fig:J1_uv2}
\end{subfigure}
\caption{Results for the $\hat{\mathcal J_1} (u,a/\sqrt{8t})$ function and continuum extrapolation.}
\label{fig:J1o2}
\end{figure}

\begin{figure}
  \centering
\end{figure}

Once more, it is convenient to study the quantity
\begin{equation}
\label{eq:comb_J1}
  \frac{1}{\hat{\mathcal J}_1(u, a/\sqrt{8t})} - \frac{1}{u} =  f_1(u, a/\sqrt{8t} )\,.
\end{equation}
The crucial difference with the previous case is that the
determination of $\mathcal J_1$ involves a change in renormalization
scale $\mu \to \mu/2$, so we expect significant scaling violations. 
On the other hand, its determination does not require to double the
lattice sizes. 
In practice we have a range in lattice spacing that spans a factor 2
further. 

Figure~\ref{fig:plot_J1} shows a comparison of the raw data with the
continuum curve (see section~\ref{sec:high-energy-regime} for a
discussion on its determination). 
In contrast with the case of $\hat{\mathcal J}_2$, we observe significant
scaling violations, confirming our hypothesis that such violations
are mainly a result of changes in the renormalization scale. Moreover,
they show a complicated functional form: the three
coarser lattices with $\sqrt{8t}/a = 1.6, 2.0, 2.4$ do not show a
monotonous pattern. 
The data is several standard deviations away from the
continuum curve. 
Figure~\ref{fig:J1_uv2} shows again the continuum extrapolation of
$\mathcal J_1(u)$ at the fixed value $u=1.5$. The plot confirms that
scaling violations are significant, with the different discretizations
based on the Wilson/Zeuthen flow showing differences of several standard deviations
for $\sqrt{8t}/a < 3.2$. 

Still, one can obtain an accurate extrapolation of $\mathcal J_1$. 
In order to do so, the large range of lattice spacings 
available to us, from $L/a=8$ to $L/a=48$, is crucial.
%in total our lattice spacing change by a factor six. 
This is of course possible only because the determination of $\mathcal
J_1$ \emph{does not require to double the lattice size}. 
Figure~\ref{fig:plot_J1} shows that the two finest lattices are in
agreement with the continuum curve. 
Note however that these are very fine lattices with $\sqrt{8t}/a
\approx 6.4, 9.6$.

\bigskip
\noindent
{\em A detailed comparison with a direct determination of $\sigma_{0.2}(u)$}
\medskip

\begin{figure}
  \centering
  \includegraphics[width=\textwidth]{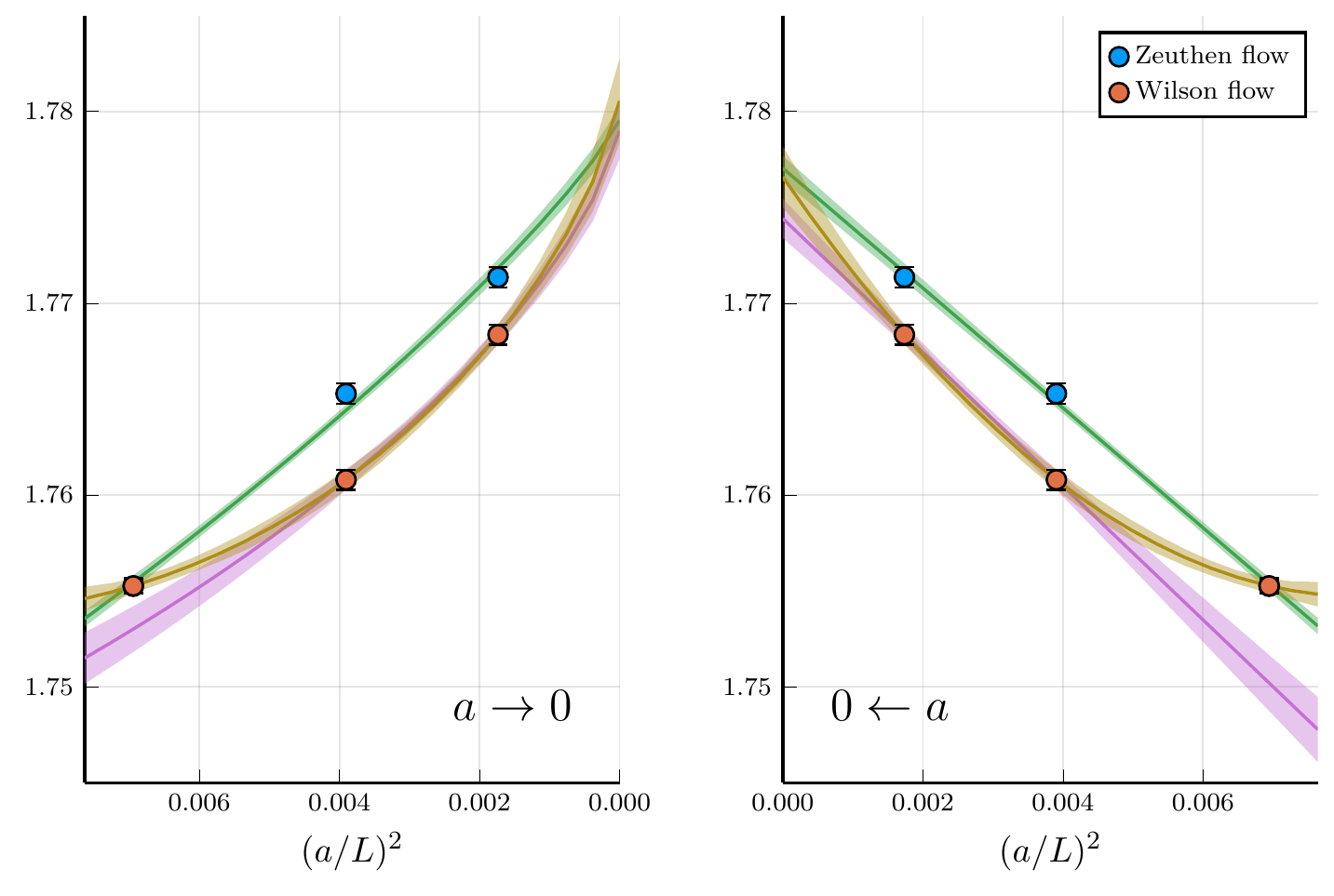} 
  \caption{Continuum extrapolation of $\Sigma_{c=0.2}(u,a/\sqrt{8t})$ for $u=1.5$.\\
  \textbf{Right:} extrapolations of the Zeuthen flow data with a
  functional form of the type $p_0 + p_1(a/L)^2$. 
  For the case of the Wilson flow one can extend the fitting range by
  including a quadratic term in the extrapolation $\mathcal O(a^4)$. 
  The extrapolated values show a reasonable agreement.\\
  \textbf{Left: } the same data is extrapolated including logarithmic
  terms in the functional ansatze. 
  For the case of the Zeuthen flow we use $p_0 +
  p_1(a/L)^2\log(a/L)$, while for the case of the Wilson flow, we
  use functional forms $p_0 + p_1(a/L)^2\log^2(a/L)$ and a three
  parameter ansatze $p_0 + p_1(a/L)^2\log^2(a/L) + (a/L)^4$ that
  allows to extend the fitting range. \\
  \textbf{Summary:} The extrapolated values vary significantly with the
  choice of logarithmic terms.
  }
  \label{fig:direc}
\end{figure}

Finally, let us compare our new strategy with the direct determination
of the step scaling function $\sigma_{c=0.2}$. 
Let's start by stating what is known\footnote{We follow the discussion
of~\cite{Husung:2019ytz}, but the interested reader should also
consult~\cite{Balog:2009np}.}. 
The leading cutoff effects of the step scaling function $\Sigma$ can
be described thanks to the Symanzik effective theory\cite{Symanzik:1970rt, Symanzik:1983dc,
  Symanzik:1981wd}. 
The asymptotic scaling violations have the form
\begin{equation}
  \Sigma(u, a/\sqrt{8t}) - \sigma(u) \sim a^2\log^{-\gamma}(a)+\dots\,.
\end{equation}
Here $\gamma$ is related with an anomalous dimension. 
Its value depends both on the details of the gauge action simulated and the
details of the observable $\Sigma$. 
Only recently~\cite{Husung:2019ytz} the leading relevant anomalous
dimension for the special case of spectral quantities (in the pure
gauge theory) has been computed. 
Except in this case, the relevant values of the leading anomalous
dimensions are unknown.

The usual linear extrapolations in $\mathcal O(a^2)$ are therefore
only justified as long as the extrapolated values do not depend
significantly on the (unknown) values of the anomalous dimensions $\gamma$. 

Figure~\ref{fig:direc} shows the extrapolation of
$\Sigma_{c=0.2}(u,a/\sqrt{8t})$ for $u=1.5$ as a representative example. 
The right panel shows linear extrapolations in $a^2$ for the Zeuthen
flow data, and both linear and quadratic extrapolations for the Wilson
flow data. The linear extrapolation of the Zeuthen flow data and the
quadratic extrapolation of the Wilson flow data show an almost perfect
agreement, with the linear extrapolation of the two finer lattice
spacings of the Wilson flow data showing also a two-sigma agreement. 

\begin{figure}
  \centering
  \includegraphics[width=\textwidth]{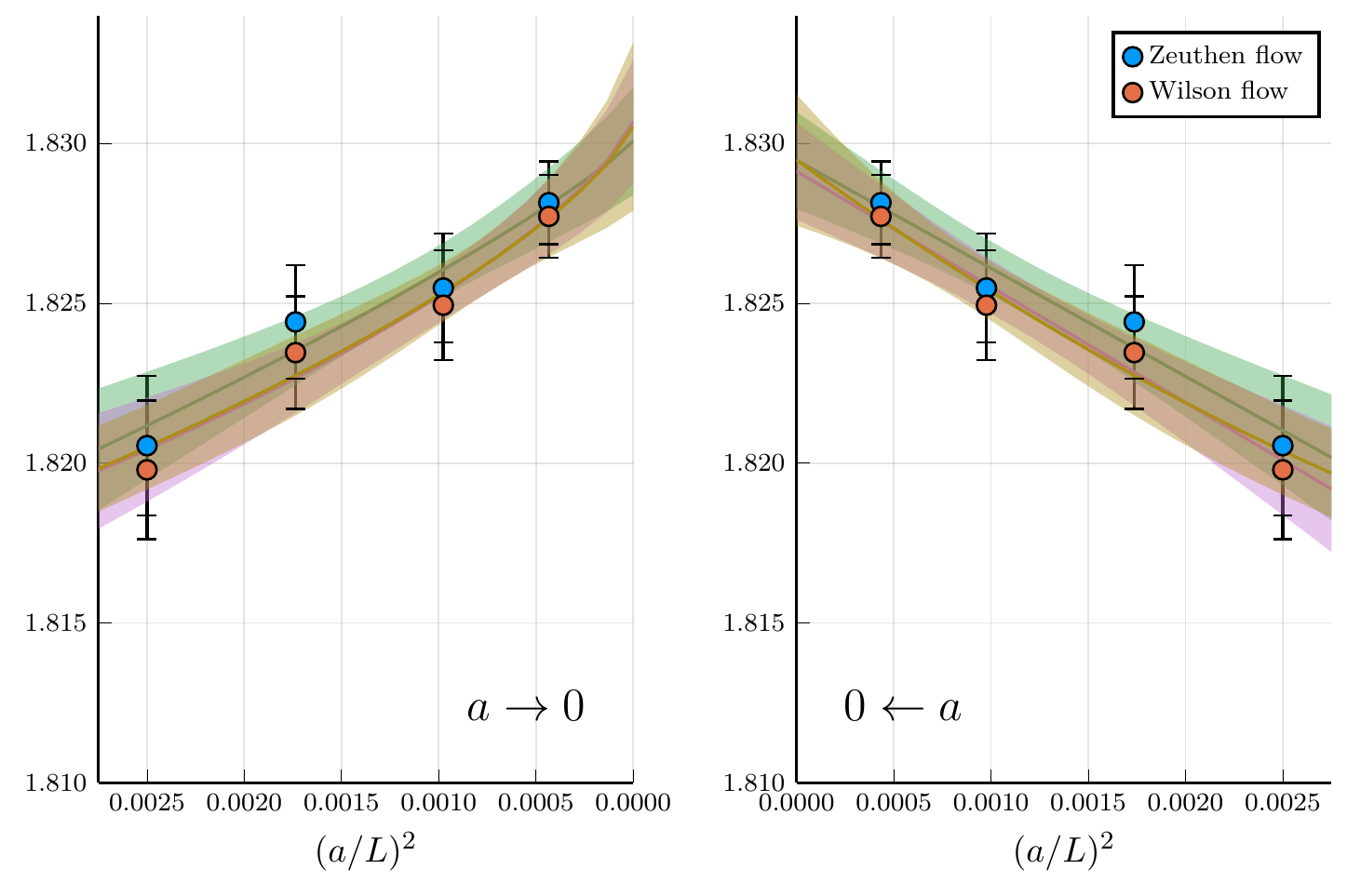} 
  \caption{Continuum extrapolation of $\mathcal J_1(u,a/\sqrt{8t})$ for $u=1.5$. 
    See text for discussion.\\
  \textbf{Right:} extrapolations of the Zeuthen flow data with a
  functional form of the type $p_0 + p_1(a/L)^2$. 
  For the case of the Wilson flow one can extend the fitting range by
  including a quadratic term in the extrapolation $\mathcal O(a^4)$. 
  The extrapolated values show a reasonable agreement.\\
  \textbf{Left: } the same data is extrapolated including logarithmic
  terms in the functional ansatze. 
  For the case of the Zeuthen flow we use $p_0 +
  p_1(a/L)^2\log(a/L)$, while for the case of the Wilson flow, we
  use functional forms $p_0 + p_1(a/L)^2\log^2(a/L)$ and a three
  parameter ansatze $p_0 + p_1(a/L)^2\log^2(a/L) + (a/L)^4$ that
  allows to extend the fitting range. \\
  \textbf{Summary:} In contrast with the case of the direct extrapolation of $\Sigma$
  (see figure~\ref{fig:direc}), all explored choices of functional
  form show a very good agreement in the extrapolated values.
  }
  \label{fig:j1details}
\end{figure}

But this consistent picture is just hiding the assumptions that are
behind such extrapolations. 
In particular, the left panel shows that using linear extrapolations in
$a^2\log(a)$ (for the Zeuthen flow data) or linear extrapolations in
$a^2\log^2(a)$ (for the Wilson flow data), one obtains an even better
agreement.  
Unfortunately the perfectly consistent extrapolations
without logarithmic terms \emph{do not agree} with the perfectly
consistent extrapolations that include such terms. 
The largest difference is between the Zeuthen flow extrapolation
in $a^2\log(a)$ (with result      1.78341(88)), and the two-point
linear extrapolation in $a^2$ of the Wilson flow data (with result
1.7744(10)), that differ by approximatley 7 combined sigmas. 
This just shows that the direct extrapolation of $\Sigma_{c=0.2}$,
although statistically very precise, has an uncontrolled systematic
uncertainty unless very large lattices are simulated. 

Figure~\ref{fig:j1details} shows the equivalent extrapolation for the
case of $\mathcal J_1$\footnote{The reader should remember that it was precisely $\mathcal J_1$ the
most challenging continuum extrapolation in our approach.}. 
In this case one can see that the extrapolations with or without
logarithmic terms agree nicely. The largest deviation is found in the
extrapolation of the Zeuthen flow data with term $a^2\log(a)$ (with
result       1.8309(19)), and the $a^2$
linear extrapolation of the Wilson flow data (with result       1.8291(15)). 
This discrepancy is less than one combined sigma.
We conclude that the uncertainty in $\mathcal J_1(u)$ is in fact
dominated by the statistical uncertainty, and not by our prejudice on
the unknown values of the anomalous dimensions. 

%%% Local Variables:
%%% mode: latex
%%% TeX-master: "paper"
%%% End:

\section{Statistical uncertainties}
\label{sec:stat-uncert}

\begin{figure}
  \centering
  \includegraphics[width=\textwidth]{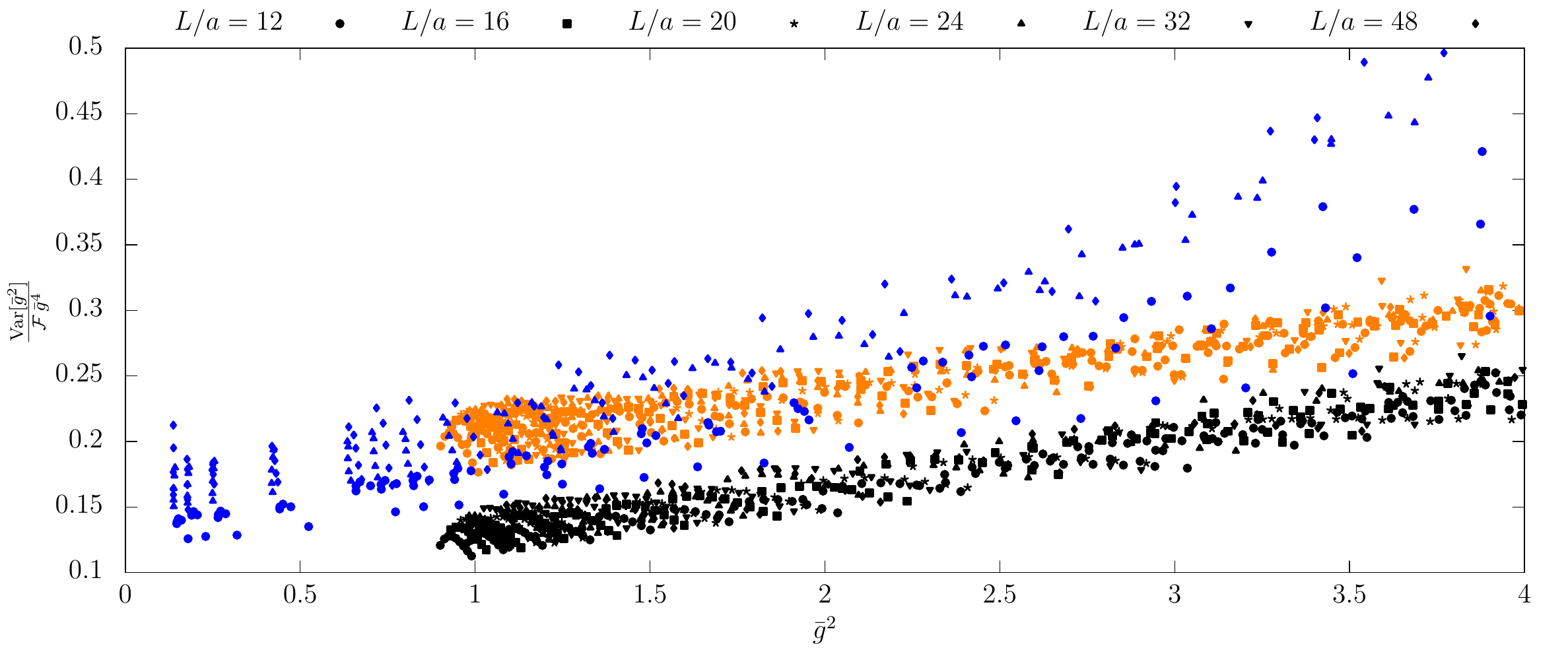} 
  \caption{Results for the quantity of eq.~(\ref{eq:var_scaling}) for three different 
  datasets and six different lattice spacings. Orange symbols are for the ``magnetic'' 
  definition of the GF coupling; black symbols are for the average of ``magnetic'' 
  and ``electric'' definitions of the GF coupling; the blue symbols are from the 
  dataset of ref.~\cite{Bribian:2020xfs}.}
  \label{fig:variance}
\end{figure}

It has been argued that a simple way to improve the scaling
properties of GF couplings consists in using large values of $c$ (see
for example the discussion in~\cite{Lin:2015zpa}). 
Unfortunately, it is well known that this comes at the cost of
increased statistical uncertainties. 
In this section we want to make this last statement more precise. 
We will present a simple model for the understanding of the scaling of the
statistical uncertainties of the GF coupling and then we will show 
how the results of numerical simulations agree with this
naive approach.

\newpage
\begin{wrapfigure}{r}{0.4\textwidth}
  \centering
  \includegraphics[width=0.35\textwidth]{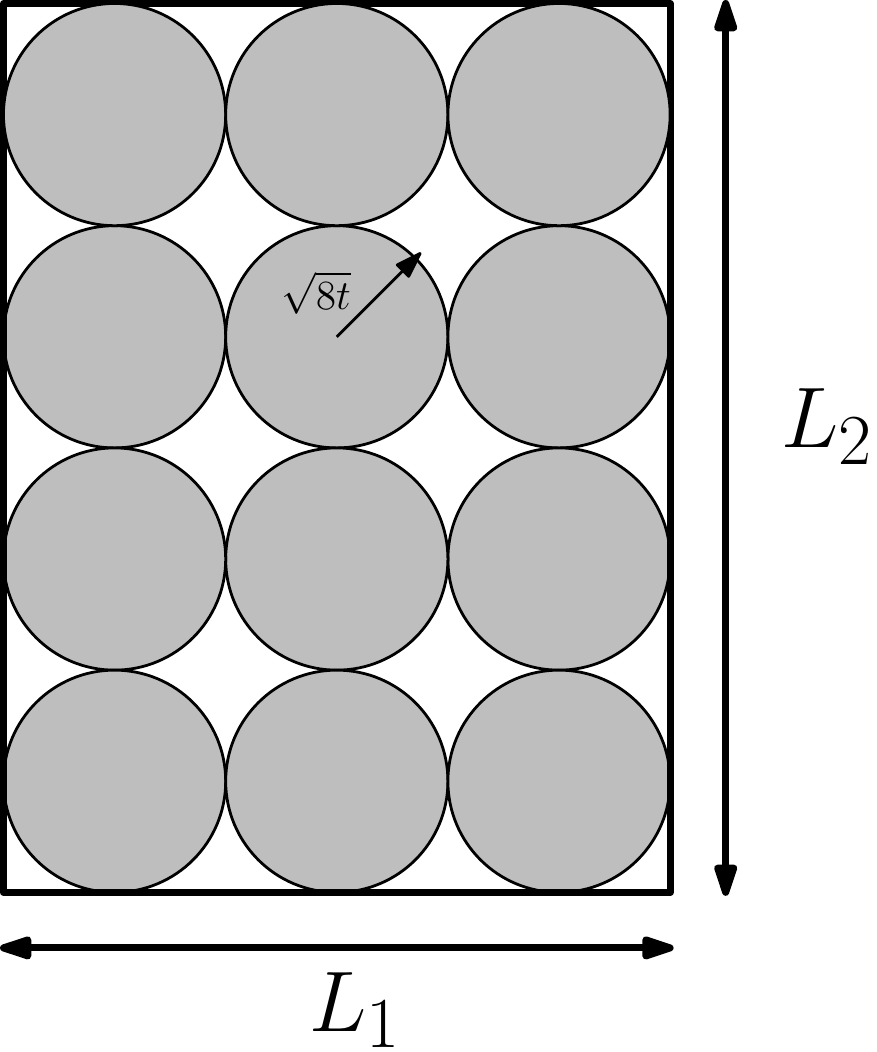}
  \caption{A simple model to explain the scaling of the variance of the GF coupling.}
  \label{fig:var_fig}
\end{wrapfigure}
Let us first focus our discussion on schemes that fully preserve the
translational invariance, like the case of periodic~\cite{Fodor:2012td}
or twisted~\cite{Ramos:2013gda} boundary conditions. 
The gradient flow smears the original gauge field $A_\mu(x)$ over a
distance $d\sim \sqrt{8t}$. 
Due to the invariance under translations, each four dimensional ball 
of radius $\sqrt{8t}$ provides an  estimate of the quantity 
$\langle E(x,t) \rangle$ (see figure~\ref{fig:var_fig}). 
Under this assumption the volume average on a lattice $L_0\times
L_1\times L_2\times L_3$ will make the variance of
the observable $\langle E(x,t) \rangle$ proportional to 
\begin{equation}
  \label{eq:ffactor}
  \mathcal F = \prod_{\mu = 0}^3 \frac{\sqrt{8t} }{L_\mu} \,.
\end{equation}
Note that in the common situation of an $L^4$ lattice with the
same length in all directions one has $\mathcal F = c^4$ (see
equation~(\ref{eq:cdef})). This gives a quantitative explanation to
the fact that the statistical uncertainties are large at large values
of $c$.

In schemes where the invariance under translations is broken in the
time direction, like Schr\"odinger Functional
(SF)~\cite{Fritzsch:2013je} or open-SF~\cite{Luscher:2014kea} boundary
conditions on a box of sizes $L_0\times L_1\times L_2\times L_3$, a
similar argument applies, except that in these cases the coupling is
only measured at a single time-slice $x_0=L_0/2$. Therefore in this case
we expect a factor 
\begin{equation}
  \label{eq:ffactor_sf}
  \mathcal F = \prod_{\mu = 1}^3 \frac{\sqrt{8t} }{L_\mu} \,,
\end{equation}
e.g. on a symmetric lattice $\mathcal F = c^3$. When do we expect this
model to break down? For the volume average 
argument to make sense, the region that is smeared by the flow must be
much smaller than the size of the lattice, so we require
\begin{equation}
\label{eq:csmall}
  \frac{\sqrt{8t} }{L_\mu} \ll 1/2\,.
\end{equation}
Note that for the case of an $L^4$ lattice this condition just means
$c\ll 0.5$. The typical values used in the literature are $c=0.2-0.4$, so
we can only expect the scaling of the variance to be approximate. 
In order to see how good this approximation is, it is useful to have a
look at the quantity
\begin{equation}
  \label{eq:var_scaling}
  \frac{{\rm Var}[\bar g ^2]}{\mathcal F \bar g ^4}\approx K(\bar g ^2)\,.
\end{equation}
If our hypothesis is correct, we expect this combination to be
independent on the lattice size and on the values of $\sqrt{8t} /L_\mu$. 
Figure~\ref{fig:variance} shows this quantity in three data sets. 

First, in orange we plot the usual magnetic coupling definition that
we have been using for our non-perturbative study (section~\ref{sec:non-pert-study}). 
This data includes values of $c=0.200, 0.225, 0.250, \dots,0.400$ for
lattices of sizes $L/a=12,16,20,24,32,48$ at several values of the
bare coupling $\beta = 6/g_0^2$. 
Note that, despite the fact that the lattice size changes by a factor
of four, and that the values of $c$ change by a factor two, the combination
in equation~(\ref{eq:var_scaling}) shows a very mild variation in all
the range of couplings $\bar g ^2 = 1-4$. 
The plot shows some variation, but to a reasonably good approximation
we can say that $Q(\bar g ^2) \approx 0.25$. 
An even better description of the data can be obtained by using a
simple linear approximation. 

Second, in black, we have another definition of the coupling in the
same datasets (in particular the lattice sizes and values of $c$ are
the same as in the previous case). The data corresponds to the coupling
definition based on the space-time components of the Energy density (i.e. 
the average between the ``magnetic'' and the ``electric'' components). 
Despite the high correlation between the electric and magnetic
energy densities, the average shows a significant smaller variance. 
In this the combination of equation~(\ref{eq:var_scaling}) can also be
reasonably well described by a linear function. 

Finally, in blue, we have data with twisted boundary condition on an
asymmetric lattice~\cite{Bribian:2020xfs}. 
In this case the simulations are done on volumes 
$L^2\times (L/3)^2$ (see~\cite{Bribian:2019ybc} for the theoretical
motivation behind this particular geometry). 
We use data with $\sqrt{8t} / L = 0.20, 0.25,\dots,0.4$ and lattice
sizes $L/a=12, 24, 48$. 
Note that in this particular case the condition
equation~(\ref{eq:csmall}) is flagrantly violated in the two short
directions, since $3\sqrt{8t} / L = 0.6-1.2$. This may explain why this
dataset shows a much larger dispersion. 
For this case the combination in equation~(\ref{eq:var_scaling})
shows a larger dependence on details like the particular choices of $L/a$ and
$\sqrt{8t}$ and not only on $\bar g ^2$.
Still, the variation is not large taking into account the disparity of
scales (varying by several factors) involved in the data (note that
naively the variance changes by more than two orders of magnitude).

It is also worth mentioning that the variance at weak coupling is
very similar for the three datasets, differing by at most a 
factor three. 
Together with the observation that being a flow observable, the
quantity in equation~(\ref{eq:var_scaling}) has a well defined
continuum limit, we can conclude that the function in
equation~(\ref{eq:var_scaling}) is \emph{universal}. 
Details like the choice of boundary conditions or the choices of
discretizations induce relatively small scaling violations,
especially at the weakest couplings. 

%%% Local Variables:
%%% mode: latex
%%% TeX-master: "paper"
%%% End:

\section{The high energy regime of Yang-Mills revisited}
\label{sec:high-energy-regime}

As a further test on our proposal, we will re-examine the high energy
regime of Yang-Mills. 
Let us first recall the relevant points of the
work~\cite{DallaBrida:2019wur}.  
\begin{itemize}
\item The determination of the $\Lambda_{\overline{\rm MS} }$
  parameter in units of $\sqrt{8t_0}$ is divided in two fundamental pieces. 
  First, a high energy part where contact with perturbation theory is made. 
  This results in a determination of 
  $\Lambda_{\overline{\rm MS} }/\mu_{\rm ref}$, with $\mu_{\rm ref}$
  being defined by $\bar g _{c=0.3}^2(\mu_{\rm ref}) =
  0.8\pi$. 
  Second, a low energy part where the dimensionless ratio $\mu_{\rm
    ref}\times \sqrt{8t_0} $ is determined.
  
\item Most of the error in the dimensionless ratio
  \begin{equation}
    \Lambda_{\overline{\rm MS} }\times \sqrt{8t_0} = \frac{\Lambda_{\overline{\rm MS} }}{\mu_{\rm ref}} \times (\mu_{\rm ref}\sqrt{8t_0})\,,
  \end{equation}
   comes from the first piece (i.e. 
   the high energy part). The total uncertainty in
   $\Lambda_{\overline{\rm MS} }\times \sqrt{8t_0}$ is 1.57\%, while
   the uncertainty in $\Lambda_{\overline{\rm MS} }/\mu_{\rm ref}$ is already 1.37\%. 
  
\item The result for $\Lambda_{\overline{\rm MS} }\times \sqrt{8t_0} =
  0.6227(98)$
  shows a significant discrepancy with other determinations: in particular 
  the very precise determination of FlowQCD~\cite{Kitazawa:2016dsl} $\Lambda_{\overline{\rm MS} }\times
  \sqrt{8t_0} = 0.5934(38)$ lies about 3 sigma away from the 
  value of~\cite{DallaBrida:2019wur}.
\end{itemize}

Given that the pure gauge determination of $\Lambda_{\overline{\rm MS}}$ 
has to face the very same challenges as the determination of
the strong coupling $\alpha_s(M_{\rm Z})$ in QCD, we think that
revising the crucial part of the work~\cite{DallaBrida:2019wur} with the new method
proposed in this work is fully justified. We recall that our dataset
is exactly the same as the one used in~\cite{DallaBrida:2019wur} (see
section~\ref{sec:non-pert-study}).

\subsection{The continuum limit of $\mathcal J_1$ and $\mathcal J_2$}
\label{sec:cont-limit-mathc}

In order to obtain the functions $\mathcal J_1$ and $\mathcal J_2$ in
the continuum, the best strategy consists in combining the continuum
extrapolation with a parametrization of the function in the continuum. 
In particular we are going to use the parametrization
\begin{equation}
  \label{eq:global}
  \frac{1}{\hat{\mathcal J}_i(u, a/\sqrt{8t})} - \frac{1}{u} =
  \sum_{n=0}^{n_{\rm c}} c_n^{(i)} u^n + \left( \frac{a}{\sqrt{8t}} \right)^2\, \sum_{n=0}^{n_\rho} \rho_n^{(i)} u^n\,.
\end{equation}
Note that the coefficients $c_n^{(i)}$ parametrize the continuum
function $\mathcal J_i(u)$, while the coefficients $\rho_n^{(i)}$
parametrize the $\mathcal O(a^2)$ cutoff effects in the function
$\hat{\mathcal J}_i$. There are several assumptions hidden in this
parametrization. 
First we assume that the continuum function $1/\hat{\mathcal J}_i(u,0) - 1/u$ can be
well described by a polynomial. 
This is certainly the case in perturbation theory to all orders\footnote{In fact 
 the first two coefficients $c_{0,1}^{(i)}$ can
 be computed with the help of the $k_i$ coefficients calculated in~\cite{DallaBrida:2017tru}:
 \begin{eqnarray*}
   c_0^{(1)} = 0.1426(7)\,, &\quad&   c_1^{(1)} = -0.0063(6)\,.\\
   c_0^{(2)} = -0.0461(7)\,, &\quad&   c_1^{(2)} = 0.0345(7)\,.
 \end{eqnarray*}}, and we expect that the non-perturbative functions 
can be well described by a polynomial ansatz. 

A more delicate assumption is that in our functional form of 
equation~(\ref{eq:global}) all scaling violations are
quadratic (i.e. $a^2/(8t)$). 
First, due to the breaking of translational invariance in the
Schr\"odinger Functional, we expect cutoff effects linear in the
lattice spacing. 
They are expected to be small, due to the localization of the GF
coupling at the timeslice $x_0=T/2$. 
Moreover the extrapolations in section~\ref{sec:non-pert-study} have
completely ignored these effects, and our data in fact seem to scale
like $\mathcal O(a^2)$ after dropping the coarser lattices. 
But due to the high precision of our data, these $\mathcal O(a)$ effects 
cannot be completely ignored, especially if we take into account 
the fact that our strategy uses data
at large values of $\sqrt{8t}/T = 0.4$, where these effects are
expected to be larger than in~\cite{DallaBrida:2019wur}, where
$\sqrt{8t}/T = 0.3$ was used. 
For this reason we include a generous estimate of these linear effects
in the error. The details are explained in Appendix~\ref{sec:boundary-mathcal-oa}. 

Our data set has also higher order cutoff effects, of the form
$a^n$ for $n>2$, and logarithmic corrections as well~\cite{Husung:2019ytz}. 
The effect of these terms in our extrapolations will be estimated by
changing the cuts used to fit the coefficients $c_n^{(i)},
\rho_n^{(i)}$. 

\subsubsection{The case $\mathcal J_1$}
\label{sec:case-mathcal-j_1}

In the exploratory study in section~\ref{sec:non-pert-study} the Zeuthen flow/improved
observable discretization already displayed better scaling properties, but still
we had to discard all lattices with $L/a< 12$. 
Since the Wilson flow/Clover combination would require even more stringent
cuts, we will only use the improved setup to quote final results. 

\begin{table}
  \centering
  \small
\begin{tabular}{llccccc}
  \toprule
  && \multicolumn{5}{c}{$u$} \\
  \cmidrule(l){3-7}
&Parameters  & 1.125 & 1.250 &  1.500 & 1.750 & 2.000 \\
\midrule
\multirow{2}{*}{$L/a\ge 12$} &
$n_{\rm c}=3,\, n_{\rho}=2$ & 1.30681(46) & 1.47575(54) & 1.82918(82) & 2.2044(12) & 2.6029(17) \\
&$n_{\rm c}=4,\, n_{\rho}=2$ & 1.30681(46) & 1.47575(54) & 1.82918(82) & 2.2044(13) & 2.6029(18) \\
\midrule
\multirow{2}{*}{$L/a\ge 16$} &$n_{\rm c}=3,\, n_{\rho}=2$ & 1.30658(49) & 1.47549(54) & 1.82882(85) & 2.2038(13) & 2.6020(19) \\
&$n_{\rm c}=4,\, n_{\rho}=2$ & 1.30658(49) & 1.47547(55) & 1.82880(85) & 2.2039(14) & 2.6021(20) \\
\midrule
\multirow{4}{*}{$L/a\ge 20$} &$\mathbf{n_{\rm c}=3,\, n_{\rho}=2}$ & \textbf{1.30640(59)} & \textbf{1.47535(61)} & \textbf{1.82878(96)} & \textbf{2.2039(16)} & \textbf{2.6021(23)} \\
&$n_{\rm c}=3,\, n_{\rho}=3$ & 1.30637(59) & 1.47550(67) & 1.8291(12)  & 2.2040(16) & 2.6014(25)  \\
&$n_{\rm c}=3,\, n_{\rho}=4$ & 1.30633(59) & 1.47549(66) & 1.8291(12)  & 2.2040(16) & 2.6014(25)  \\
&$n_{\rm c}=4,\, n_{\rho}=2$ & 1.30639(59) & 1.47539(62) & 1.82881(96) & 2.2038(16) & 2.6019(23) \\
\bottomrule
\end{tabular}
  \caption{Values of the continuum function $\mathcal J_1(u)$ for
    different fit parametrizations and cuts (see equation~(\ref{eq:global})). 
    In bold we show our preferred fit. 
    See text for more details.
  }
  \label{tab:J1}
\end{table}

This is confirmed by looking at table~\ref{tab:J1}, where the
values in the continuum of $\mathcal J_1(u,0)$ are shown at a few
representative values of $u$. 
As the reader can see, the effect of varying the number of fit
parameters ($n_{\rm c}$ and $n_\rho$ in equation~(\ref{eq:global})) is
negligible. On the other hand, the cut in $L/a$ has a small effect on
the extrapolations. 
If lattices with $L/a=12$ are included, the continuum value of
$\mathcal J_1$ seems to be systematically higher, but still compatible
within errors. 
Also the statistical errors are smaller for these analysis. 
A conservative approach consists in just taking a fit with $L/a\ge 20$
(i.e.  $a/\sqrt{8t} < 1/4$), so that the continuum value has a larger uncertainty. 
Note that since the computation of $\mathcal J_1$ does not require to
double the lattice sizes, even with this stringent cut our dataset
still offers more than a factor two in lattice spacing. 
Among these fits there is very little difference
between different parametrizations. 
Moreover the fit quality is very similar in all cases. 
All in all we just choose one of these fits ($n_{\rm c}=3$,
$n_\rho=2$, bold in table~\ref{tab:J1}) as our final result. 

\subsubsection{The case $\mathcal J_2$}
\label{sec:case-mathcal-j_2}

\begin{table}
  \centering
  \small
  \begin{tabular}{llccccc}
  \toprule
  && \multicolumn{5}{c}{$u$} \\
  \cmidrule(l){3-7}
&Parameters  & 1.125 & 1.250 &  1.500 & 1.750 & 2.250 \\
\midrule
\multirow{2}{*}{All $L/a$} &
$n_{\rm c} =3, n_\rho=2$ & 1.09602(61) & 1.21647(67) & 1.45772(97) & 1.6999(14) & 2.1879(23) \\
&$n_{\rm c} =4, n_\rho=2$ & 1.09607(62) & 1.21640(67) & 1.45764(96) & 1.7000(14) & 2.1879(23) \\
\midrule
\multirow{2}{*}{$L/a\ge 10$} &
$n_{\rm c} =3, n_\rho=2$ & 1.09640(68) & 1.21677(70) & 1.45778(96) & 1.6996(14) & 2.1871(24) \\
&$n_{\rm c} =4, n_\rho=2$ & 1.09647(69) & 1.21672(70) & 1.45770(96) & 1.6997(14) & 2.1871(24) \\
\midrule
    \multirow{4}{*}{$L/a\ge 12$} &
$\mathbf{n_{\rm c} =3, n_\rho=2}$ & \textbf{1.09629(82)} & \textbf{1.21652(81)} & \textbf{1.4572(10)} & \textbf{1.6986(16)} & \textbf{2.1859(28)} \\
&$n_{\rm c} =3, n_\rho=3$ & 1.09621(89) & 1.21658(82) & 1.4574(13) & 1.6988(17) & 2.1852(35) \\
&$n_{\rm c} =3, n_\rho=4$ & 1.09629(89) & 1.21672(83) & 1.4575(13) & 1.6985(17) & 2.1843(36) \\
&$n_{\rm c} =4, n_\rho=2$ & 1.09650(85) & 1.21641(81) & 1.4569(11) & 1.6986(16) & 2.1858(28) \\
\bottomrule
\end{tabular}
  \caption{Values of the continuum function $\mathcal J_2(u)$ for
    different fit parametrizations and cuts (see equation~(\ref{eq:global})). 
    In bold we show our preferred fit. 
    See text for more details.
  }
  \label{tab:J2}
\end{table}

The computation of $\mathcal J_2$ requires to double the lattice
sizes, and then our datasets offers only half the lever arm in lattice
spacing for the continuum extrapolations. 
Our hypothesis is that the scaling violations are small for $\mathcal
J_2$ because its determination does not involve a change in
renormalization scale. Our preliminary investigation of
section~\ref{sec:non-pert-study} has also confirmed this hypothesis. 
Table~\ref{tab:J2} shows that this is indeed the case. 
Even including the coarser lattices with $L/a=8$ (corresponding to
$a/\sqrt{8t} = 1/1.6$), the results are in agreement within errors. 
It is clear that the choice of parametrization has very little effect. 
We just settle for one particular fit with $L/a\ge 12$ (represented in
bold in table~\ref{tab:J2}) that we will use for any further analysis.

\subsection{The quantity $\sqrt{8t_0}  \times
  \Lambda_{\overline{\rm MS} }$}
\label{sec:quantity-sq8t-times}

\subsubsection{The scale $\mu_{\rm ref}$}
\label{sec:scale-mu_ref}

As we have already mentioned, the original
work~\cite{DallaBrida:2019wur} determined the dimensionless
combination $\sqrt{8t_0}  \times \Lambda_{\overline{\rm MS} }$ as the
product of two factors. 
First the low energy factor
\begin{equation}
  \label{eq:t0muref}  
  \sqrt{8t_0}\mu_{\rm ref} =        7.808(46) \qquad    [0.59\%]\,,
\end{equation}
that has a very small uncertainty\footnote{The quantity
  $\sqrt{8t_0}\mu_{\rm ref}$ was determined in two different schemes,
  and for each scheme, using two different strategies. 
All analysis resulted in completely negligible differences. 
}. The other factor $\Lambda_{\overline{\rm MS} }/\mu_{\rm ref}$ was
much more delicate to determine. 
It is precisely this last quantity that we want to determine once more
with our new strategy. 
A first step consists in dealing with the factor $\mu_{\rm ref}$. This
was defined in the scheme with $c=0.3$ by the condition
\begin{equation}
  \label{eq:gmuref_m}
  \bar g^2_{c=0.3}(\mu_{\rm ref}) = \frac{4\pi}{5}
  \approx 2.5132\ldots\,.
\end{equation}
Since our new strategy provides the step scaling function
$\sigma(u)$ for $c=0.2$, we must first
determine the value of our coupling $\bar g _{c=0.2}(\mu_{\rm ref})$. 
The procedure is completely analogous to the determination of
$\mathcal J_1$. 
We first define
\begin{equation}
\label{eq:J3def}
  \hat{\mathcal J}_3 (u,a/\sqrt{8t}) = \bar g ^2_{c=0.3}(2\mu/3)\Big|_{\bar g ^2_{c=0.2}(\mu) = u}\,.
\end{equation}
We choose to fit our data to the model
\begin{equation}
  \frac{1}{\hat{\mathcal J}_3 (u,a/\sqrt{8t})} - \frac{1}{u} = \sum_{n=0}^{n_{\rm c}}c_n^{(3)}u^n
  + \left( \frac{a}{\sqrt{8t}} \right)^2\, \sum_{n=0}^{n_\rho} \rho_n^{(i)} u^u\,.
\end{equation}
The same considerations discussed in section~\ref{sec:case-mathcal-j_1}
apply to the determination, in the continuum, of the relation between
$\bar g^2_{c=0.2}(\mu)$ and $\bar g^2_{c=0.3}(\mu)$. 
In this case, however, we expect the scaling violations to be smaller,
since the change in renormalization scale is not a factor two, but
only a factor 3/2.

We performed several fits, changing the number of fit parameters
$n_{\rm c}$ and $n_\rho$, and using different cuts for our data, and
the overall analysis results in a consistent value for 
$\bar g_{c=0.2}(\mu_{\rm ref})$ as long as data with $L/a\ge 16$ is used.

\begin{figure}
  \centering
  \includegraphics[width=\textwidth]{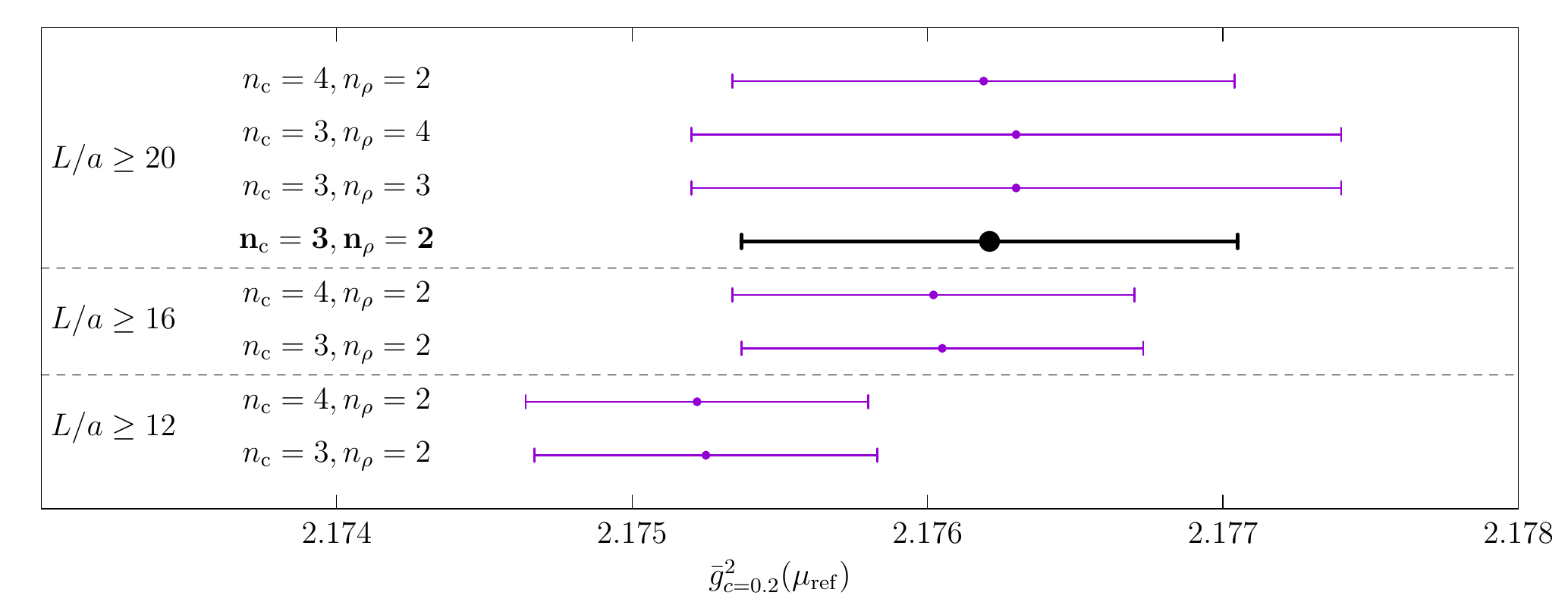} 
  \caption{Determination of $\bar g ^2_{c=0.2}(\mu_{\rm ref})$ using
    different parametrizations and cuts. In black and bold face the
    result of equation~(\ref{eq:muref})).
  }
  \label{fig:muref}
\end{figure}

We choose to quote the result with $n_{\rm c}=3$ and $n_\rho=2$ and
$L/a\ge 20$
\begin{equation}
  \label{eq:muref}
  \bar g ^2_{c=0.2}(\mu_{\rm ref}) =      2.17621(84)\,.
\end{equation}
Despite the high precision, the result should be actually considered conservative,
as this particular fit has one of the largest uncertainties of all the
combinations that we tried (see figure~\ref{fig:muref}).  

\subsubsection{The extraction of $\Lambda_{\overline{\rm MS} }$}
\label{sec:extraction-lambda}

The $\Lambda_s$-parameter in the scheme defined by the coupling $\bar
g _s^2$ is given by the expression
\begin{equation}
  \label{eq:lamdef}
  \frac{\Lambda_s}{\mu} = 
  \left[b_0\bar g_s^2(\mu)\right]^{-\frac{b_1}{2b_0^2}}\,
  e^{-\frac{1}{2b_0\bar g_s^2(\mu)}}\,
  \exp\{-I_s(\bar{g_s}(\mu))\},
  \quad
  I_s(g)= \int_{0}^{g}{\rm d}x\, \left[\frac{1}{\beta_s(x)} + 
  \frac{1}{b_0x^3} - \frac{b_1}{b_0^2x}\right]\,.
\end{equation}
Note that this expression is exact, and valid beyond perturbation theory, 
as long as the non-perturbative $\beta$-function, defined by
\begin{equation}
  \label{eq:beta}
  \mu\frac{{\rm d}}{{\rm d}\mu} \bar g_s(\mu) = \beta_s(\bar g)\,.
\end{equation}
is known. If two renormalized couplings are related to one-loop by the
expression   
\begin{equation}
  \label{eq:gssp}
  \bar g^2_{s'}(\mu) = \bar g^2_s(\mu)  +   c_{ss'}\bar g^4_s(\mu) +\dots
\end{equation}
the corresponding $\Lambda$-parameters are related by
\begin{equation}
  \label{eq:lam_change}
  \frac{\Lambda_{s'}}{\Lambda_s} = 
  \exp\left(\frac{-c_{ss'}}{2b_0}\right)\,.
\end{equation}
This last formula allows for a non-perturbative definition of
$\Lambda_{\overline{\rm MS} }$, even if the $\overline{\rm MS} $ 
scheme is intrinsically perturbative. 

All in all, the determination of $\Lambda_{\overline{\rm MS} }$
requires the determination of the integral in equation~(\ref{eq:lamdef}) 
in a scheme that is non-perturbatively defined. 
The lower limit of the integral is zero, which requires
to determine the $\beta$-function up to infinite energy. 
In practice this can only be achieved by a limit process. 
One first defines
\begin{equation}
  \label{eq:JPT}
  K_s(\bar g _s(\mu),g_{\rm PT})= \int_{g_{\rm PT}}^{\bar g _s(\mu)}{\rm d}x\, \left[\frac{1}{\beta_s(x)} + 
  \frac{1}{b_0x^3} - \frac{b_1}{b_0^2x}\right] + \int_0^{g_{\rm PT}}{\rm d}x\, \left[\frac{1}{\beta_s^{(l)}(x)} + 
  \frac{1}{b_0x^3} - \frac{b_1}{b_0^2x}\right]\,,
\end{equation}
very similar to the previous function  $I_s(\bar g _s(\mu))$. 
The only difference is that the integral in
equation~(\ref{eq:lamdef}) for values of the coupling smaller than 
$g_{\rm PT}$ is determined by substituting the 
$\beta_s$-function by its $l$-loop perturbative
approximation $\beta_s^{(l)}(x)$
\begin{equation}
  \beta_s(x) \simas{x\to0} \beta_s^{(l)}(x) = -x^3\sum_{n=0}^l b_n x^{2n} + \mathcal O(x^{2(l+1)})\,.
\end{equation}
The first two coefficients $b_0 = 11/(4 \pi)^2$ and $b_1 = 102/(4 \pi)^4$ 
are scheme-independent, while the values $b_n$ for $n>1$ depend on the chosen scheme. 
It is now clear that 
\begin{equation}
  \label{eq:IPT}
  K_s(\bar g _s(\mu),g_{\rm PT}) \simas{g_{\rm PT}\to 0} I_s(\bar g _s(\mu)) +  \mathcal O(g_{\rm PT}^{2(l-1)})\,,
\end{equation}
In practice for most finite volume schemes, the $\beta_s$-function is
known up to three loops, and therefore the corrections are $\mathcal
O(g_{\rm PT}^4)$. The value of the coupling $g_{\rm PT}$ delimits the
energy region (from $\mu_{\rm PT}$ to $\infty$) where perturbation
theory is used via the relation 
\begin{equation}
  \bar g_s^2(\mu_{\rm PT}) = g_{\rm PT}^2\,.
\end{equation}
Ideally one would like to estimate the $\Lambda$-parameter by taking
the following limit
\begin{equation}
\label{eq:lamlimit}
  \frac{\Lambda_s}{\mu} = \lim_{g_{\rm PT}\to 0} \left\{ 
  \left[b_0\bar g_s^2(\mu)\right]^{-\frac{b_1}{2b_0^2}}\,
  e^{-\frac{1}{2b_0\bar g_s^2(\mu)}}\,
  \exp\{-K_s(\bar{g_s}(\mu),g_{\rm PT})\}
   \right\} \,.
\end{equation}
Since the value of the coupling $\bar g^2_s(\mu)$ runs
logarithmically with $\mu$, it is technically a challenge to probe a
large range of energy scales so that the corrections $\mathcal O(g
^{2(l-1)}_{\rm PT})$ vary substantially and the limit can be taken accurately. 

Of course finite-size scaling \emph{was designed} to explore such
large ranges of energy scales. 
Starting from the scale $\mu_{\rm ref}$ (see
section~\ref{sec:scale-mu_ref}), and with the knowledge of the step
scaling function $\sigma = \mathcal J_2\circ \mathcal J_1$
(see section~\ref{sec:cont-limit-mathc}), one can define the sequence
of couplings
\begin{equation}
\label{eq:gseq}
  u_0 = \bar g ^2(\mu_{\rm ref})\,, \qquad
  u_n = \sigma^{-1}(u_{n-1}) = \mathcal J_1^{-1} \left(\mathcal J_2^{-1}(u_{n-1})  \right) =
  \bar g ^2(2^n\mu_{\rm ref})\,.
\end{equation}
The energy scales reached by this procedure increase geometrically. 
Contact with perturbation theory can be made at each step by choosing
$g^2_{\rm PT} = u_n$ in equation~(\ref{eq:IPT}), and one can indeed
check that the corrections $\mathcal O(u_n^2)$ are small and decrease
as they should. 
For a long time, the challenge was mainly to maintain
a high precision, but the most recent works~\cite{DallaBrida:2018rfy,
  DallaBrida:2019wur} have shown that when one reaches a high
precision, the corrections can be significant in some schemes even at very
high energy scales.  

\begin{table}
  \centering
\footnotesize
%\small
  \begin{tabular}{lcccccccc}
  \toprule
  & \multicolumn{2}{c}{GF} & \multicolumn{2}{c}{SF} & \multicolumn{2}{c}{$\overline{\rm MS}$ $(s=1)$} & \multicolumn{2}{c}{$\overline{\rm MS}$ $(s=2)$} \\
\cmidrule(l){2-3} \cmidrule(l){4-5} \cmidrule(l){6-7} \cmidrule(l){8-9}
  $n$& $\bar g^2 (\mu_n)$ & $\sqrt{8t_0}\, \Lambda_{\overline{\rm MS} }$
                            & $\bar g^2_{\rm SF} (0.3\mu_{n+1})$ & $\sqrt{8t_0}\, \Lambda_{\overline{\rm MS} }$
                            & $\bar g^2_{\overline{\rm MS} } (0.3\mu_{n+1})$ & $\sqrt{8t_0}\, \Lambda_{\overline{\rm MS} }$
                            & $\bar g^2_{\overline{\rm MS} } (0.3\mu_{n+2})$ & $\sqrt{8t_0}\, \Lambda_{\overline{\rm MS} }$\\
  \midrule
0&2.17621(84) & 0.6714(33)  & 2.008(10) & 0.6179(99) & 2.554(16) & 0.5615(85) & 2.037(11) & 0.6112(97)\\
1&1.7734(22) & 0.6596(49)   & 1.6553(24) & 0.6252(46) & 2.0087(36) & 0.5850(42) & 1.6732(25) & 0.6205(45)\\
2&1.4989(28) & 0.6527(67)   & 1.4091(25) & 0.6270(62) & 1.6566(34) & 0.5972(58) & 1.4211(25) & 0.6237(62)\\
3&1.2992(27) & 0.6479(83)   & 1.2280(24) & 0.6267(77) & 1.4112(32) & 0.6037(74) & 1.2366(25) & 0.6242(77)\\
4&1.1471(25) & 0.6442(95)   & 1.0890(23) & 0.6252(90) & 1.2302(29) & 0.6070(87) & 1.0955(23) & 0.6233(90)\\
5&1.0273(24) & 0.641(11)  & 0.9789(22) & 0.623(11) & 1.0911(28) & 0.608(10) & 0.9839(22) & 0.621(11)\\
    \midrule
$\infty$ & {Linear} & 0.631(15) &  &        0.621(16) &  &        0.618(14) &  &        0.621(16) \\
$\infty$ & {Constant} & - &  &       0.6260(74) &  & - &  &       0.6234(73) \\
    \bottomrule
\end{tabular}
  \caption{Sequence of couplings in different schemes and at different
  scales ($\mu_n = 2^n\mu_{\rm ref}$) and the corresponding values of
  $\sqrt{8t_0}\,  \Lambda_{\overline{\rm MS} }$ (see text for more details). 
The values of $\bar g (\mu_n)$ are obtained via a recursive application of the step scaling
function in the GF scheme (equation~(\ref{eq:gseq})). 
The conversion from the GF scheme to the SF scheme is performed
non-perturbatively and detailed in
appendix~\ref{sec:matching-with-sf}. 
The conversion to the $\overline{\rm MS} $ scheme is done by using the
perturbative relation with the SF scheme (equation~(\ref{eq:sftoms})). 
The last two rows show possible extrapolations of
$\sqrt{8t_0}\, \Lambda_{\overline{\rm MS} }$ using the last four
values ($n=2,3,4,5$). 
We show both an extrapolation linear in the $\bar g ^4$ and a
extrapolation to a constant for the cases that this behavior is
compatible with the data.}
  \label{tab:lam}
\end{table}

For this reason \emph{reaching high energies alone is not
  enough}. The limit in equation~(\ref{eq:lamlimit}) has to be
taken seriously and the systematics well estimated. 
Fortunately our dataset allows us to study the matching
with perturbation theory at energy scales
\begin{equation}
  \label{eq:mun}
  \mu_n = 2^n\mu_{\rm ref}\,, \qquad \left( n=0,\dots,5 \right)\,.
\end{equation}
Moreover we will explore several options to match with perturbation theory:
\begin{description}
\item[GF:] This is just the direct application of
  equation~(\ref{eq:lamdef}) using $g^2_{\rm PT} = u_n$ to determine
  $I_{\rm GF}(\bar g(\mu_n))$ (cf. 
  equation~(\ref{eq:IPT})). Schematically:
  \begin{displaymath}
    u_n = \bar g ^2(\mu_n) \xrightarrow[(\text{eq.}~(\ref{eq:lamdef}))]{\beta^{(3)}_{\rm GF}} \frac{\Lambda_{\rm GF}}{\mu_{\rm ref}} \xrightarrow{\frac{\Lambda_{\overline{\rm MS} }}{\Lambda_{\rm GF}}}
    \frac{\Lambda_{\overline{\rm MS} }}{\mu_{\rm ref}}\,.
  \end{displaymath}

  In this case the matching with perturbation theory is performed in the
  GF scheme at a scale $\mu_{\rm PT} = \mu_n = 2^n\mu_{\rm ref}$.
  
\item[SF:] Reference~\cite{DallaBrida:2019wur} showed that schemes
  based on the GF show a very poor perturbative convergence. The same
  reference suggested to match non-perturbatively to the traditional
  SF coupling~\cite{Luscher:1992an} with background field. 
  The details of this matching are explained in appendix~\ref{sec:matching-with-sf}. 
  Schematically:
  \begin{displaymath}
    u_n = \bar g ^2(\mu_n)
    \xrightarrow[(\text{ap}.~\ref{sec:matching-with-sf})]{\text{GF}\to\text{SF}} \bar g^2_{\rm SF}(0.3\,
  \mu_{n+1})
    \xrightarrow[(\text{eq.}~(\ref{eq:lamdef}))]{\beta^{(3)}_{\rm SF}} \frac{\Lambda_{\rm SF}}{\mu_{\rm ref}} \xrightarrow{\frac{\Lambda_{\overline{\rm MS} }}{\Lambda_{\rm SF}}}
    \frac{\Lambda_{\overline{\rm MS} }}{\mu_{\rm ref}}\,.
  \end{displaymath}

  In this case matching with perturbation theory is performed in the
  SF scheme at a scale $\mu_{\rm PT} = 0.3\times \mu_{n+1} = 0.3\times
  2^{n+1}\mu_{\rm ref}$.

\item[$\overline{\mbox{MS}}$:] One can convert the values of the SF
  coupling to the $\overline{\rm MS} $ scheme using the perturbative
  relation~\cite{Bode:1998hd}
  \begin{equation}
    \label{eq:sftoms}
    \bar g ^2_{\overline{\rm MS} }(s\mu) = \bar g ^2_{\rm SF}(\mu) +
    \frac{c_1(s)}{4\pi} \bar g ^4_{\rm SF}(\mu) + 
    \frac{c_2(s)}{(4\pi)^2} \bar g ^6_{\rm SF}(\mu) + \dots\,.
  \end{equation}
   where
  \begin{subequations}
\label{eq:ccoef}
    \begin{eqnarray}
   c_1(s) &=& -8\pi b_0\log s + 1.255621(2)\,,\\
   c_2(s) &=& c_1^2(s) - 32\pi^2 b_1 \log(s) + 1.197(10)\,.
  \end{eqnarray}
  \end{subequations}
  
  Once the value of the coupling in the $\overline{\rm MS} $ scheme is
  known, one can use the known 5-loop $\beta$-function~\cite{Baikov:2016tgj}
  to determine directly $\Lambda_{\overline{\rm MS} }$. 
  Even if the running is known much more accurately in the
  $\overline{\rm MS}$ scheme, this procedure carries the same
  parametric uncertainty $\mathcal O(\bar g ^4_{\overline{\rm MS} })$
  as the others, since the limiting factor is represented by the known orders in the
  perturbative relation between couplings, equation~(\ref{eq:sftoms})
  (see~\cite{DallaBrida:2018rfy}).    
  Schematically we have
  \begin{displaymath}
    u_n = \bar g ^2(\mu_n)
    \xrightarrow[(\text{ap}.~\ref{sec:matching-with-sf})]{\text{GF}\to\text{SF}} \bar g^2_{\rm SF}(
    0.3\,\mu_{n+1}) \xrightarrow[(\text{eq.~(\ref{eq:sftoms})})]{\text{SF}\to \overline{\rm MS} } \bar g^2_{\overline{\rm MS} }(s\, 0.3\,\mu_{n+1})
    \xrightarrow[(\text{eq.}~(\ref{eq:lamdef}))]{\beta^{(5)}_{\overline{\rm MS} }} 
    \frac{\Lambda_{\overline{\rm MS} }}{\mu_{\rm ref}}\,.
  \end{displaymath}

 In this case the scale of matching with perturbation theory is
 performed in the SF scheme at a scale $\mu_{\rm PT} = s\, 0.3\times
 2^{n+1}\mu_{\rm ref}$, but the RG evolution is done in the
 $\overline{\rm MS}$ scheme. The value of $s$ is in principle
 arbitrary, but if taken too large the perturbative coefficients of
 equation~(\ref{eq:ccoef}) become large, and one expects a bad
 asymptotic convergence of the perturbative series. 
 We will explore two choices: first the simple $s=1$, and then the value $s=2$, 
 that is very close to the value of \emph{fastest apparent convergence}
 \footnote{The scale of fastest
   apparent convergence is defined by a vanishing 1-loop coefficient
   in the relation between the coupling and the $\overline{\rm MS} $ scheme (i.e. 
   $c_1(s)=0$ in equation~(\ref{eq:ccoef})).}.
  
\end{description}

The values of $\Lambda_{\overline{\rm MS} }/\mu_{\rm ref}$ can be
multiplied by the factor $\sqrt{8t_0}\, \mu_{\rm ref} $ (cf. 
equation~(\ref{eq:t0muref})) to produce the results for $\sqrt{8t_0}\,
\Lambda_{\overline{\rm MS}} $ reported in table~\ref{tab:lam} according to the
different procedures. In the next section we will comment on the
results.  

\subsubsection{Results and discussion}
\label{sec:results}

We refer the reader once more to table~\ref{tab:lam}. 
The values for $\sqrt{8t_0}\, \Lambda_{\overline{\rm MS}}$ differ
for the different treatments of perturbation theory. 
There are two important points worth mentioning:
\begin{enumerate}
\item Even at scales where $\bar g^2\approx 1$ (corresponding to $\alpha\approx
  0.08$), different treatments of perturbation theory produce values
  of $\sqrt{8t_0}\, \Lambda_{\overline{\rm MS}}$ that vary as much as 3\%. 
\item There are two particular treatments of perturbation theory
  (labeled SF and $\overline{\rm MS} (s=2)$), where the value of
  $\sqrt{8t_0}\, \Lambda_{\overline{\rm MS} }$ is constant within
  errors when extracted over a range of energy scales that vary by a
  factor 32.
\end{enumerate}

\begin{figure}
  \centering
  \includegraphics[width=\textwidth]{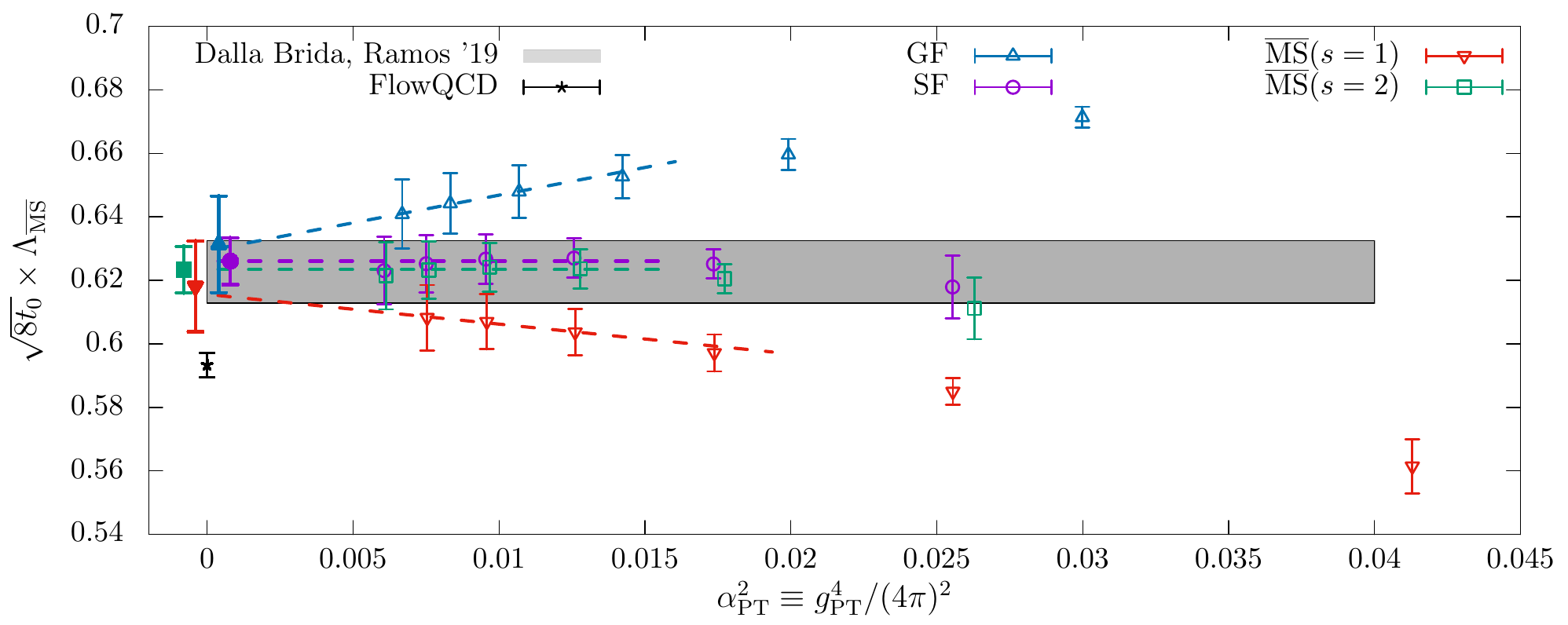} 
  \caption{The dimensionless product $\sqrt{8t_0} \times
    \Lambda_{\overline{\rm MS} }$ as a function of $g_{\rm PT}$ (see equation~(\ref{eq:lamlimit})). 
The empty symbols represent the data of table~\ref{tab:lam} for $n=0,\dots,5$, while 
the  filled symbols are extrapolations $g_{\rm PT}\to 0$ (shifted for
better visibility) of the
different approaches to the perturbative matching (see text for more
details). The gray band is the result of
reference~\cite{DallaBrida:2019wur}, while the data point labeled
FlowQCD is the result of reference~\cite{Kitazawa:2016dsl}.}
\label{fig:lam}
\end{figure} 

These results are also plotted in figure~\ref{fig:lam}. 
Qualitatively we see  that the variations between different treatments
of perturbation theory roughly scale as expected (i.e. 
decrease proportionally to $\alpha^2$). 

A more quantitative picture is obtained by looking at the two last
rows of table~\ref{tab:lam}. 
They show possible extrapolations of the quantity $\sqrt{8t_0}\,
\Lambda_{\overline{\rm MS}}$ (see equation~(\ref{eq:lamlimit})): the
deviation from the final result of reference~\cite{DallaBrida:2019wur}
$\sqrt{8t_0}\, \Lambda_{\overline{\rm MS}} = 0.6227(98)$ of any of the
possible extrapolations is below the statistical uncertainties (about
$1.5\%$).  This is half of the differences present at scales where
$\bar g^2 \approx 1$.

All extrapolations $g_ {\rm PT}\to 0$ agree well (last two rows of table~\ref{tab:lam}). 
In particular even the extrapolations that assume that the higher order
terms proportional to $g^4_{\rm PT}$ are negligible show quite a
small uncertainty. 
Still, the error band covers the central values of all other extrapolations. 
Note however that the size of the uncertainties depends strongly on
how much data one decides to include. 
A very conservative approach (such as the one used in
reference~\cite{DallaBrida:2019wur}) would consist in just
quoting as final result the value at the most perturbative point. 
This is justified since the data labeled SF and $\overline{\rm MS}\, 
(s=2)$ shows basically no dependence on the value of $g_{\rm PT}$.

Note however that the methodology in these two works is very different. 
In particular they deal with the systematic associated with the
continuum extrapolations in a very different way.
Reference~\cite{DallaBrida:2019wur} uses the GF coupling with $c=0.3$
in order to perform the non-perturbative running. 
On the other hand we use the step scaling function with $c=0.2$,
determined as the composition of the functions $\mathcal J_1$ and
$\mathcal J_2$ as described in section~\ref{sec:new-strat-determ} 
order to do the non-perturbative running. 
%Therefore we find quite comforting that the results, both the central
%values and the uncertainties, are basically identical. 
Even with the highly conservative approach that we used in 
section~\ref{sec:cont-limit-mathc} to perform the continuum 
extrapolations of $\hat{\mathcal{J}_1}$ and $\hat{\mathcal{J}_2}$,
we find a final uncertainty on $\sqrt{8t_0}\, \Lambda_{\overline{\rm MS}}$
of the same size of the one obtained in reference~\cite{DallaBrida:2019wur}.
Moreover, the fact that the central values are in perfect agreement
in the two calculations provides very strong evidence that the systematic 
effects associated with the continuum extrapolations are completely under control,
and well below our statistical uncertainties.

Finally, the extrapolations that assume a linear dependence in
$\alpha_{\rm PT}^2$ show larger uncertainties. Still, it is very
important to see that the $g_{\rm PT}\to 0$ extrapolation
substantially improves the agreement between all treatments of the perturbative matching. 

It is also worth mentioning that the propagation of the linear
$\mathcal O(a)$ effects (see appendix~\ref{sec:boundary-mathcal-oa})
represent about a 15\% of the final error squared in $\sqrt{8t_0}\,
\Lambda_{\overline{\rm MS}}$. 
This is about a 50\% larger than in the extraction of
reference~\cite{DallaBrida:2019wur} and can be understood noting that
making use of values of the coupling at $c=0.4$, we increase the
boundary effects. 

All in all our approach shows a remarkable agreement between very
different treatments of the matching with perturbation theory, and
thanks to our new proposal, we are able to also show a 
very good agreement with previous works that have rather
different systematics associated with the continuum extrapolation.

\subsubsection{Scale uncertainties}
\label{sec:scale-uncertainties}

\begin{wrapfigure}{r}{0.5\textwidth}
  \centering
  \includegraphics[width=0.49\textwidth]{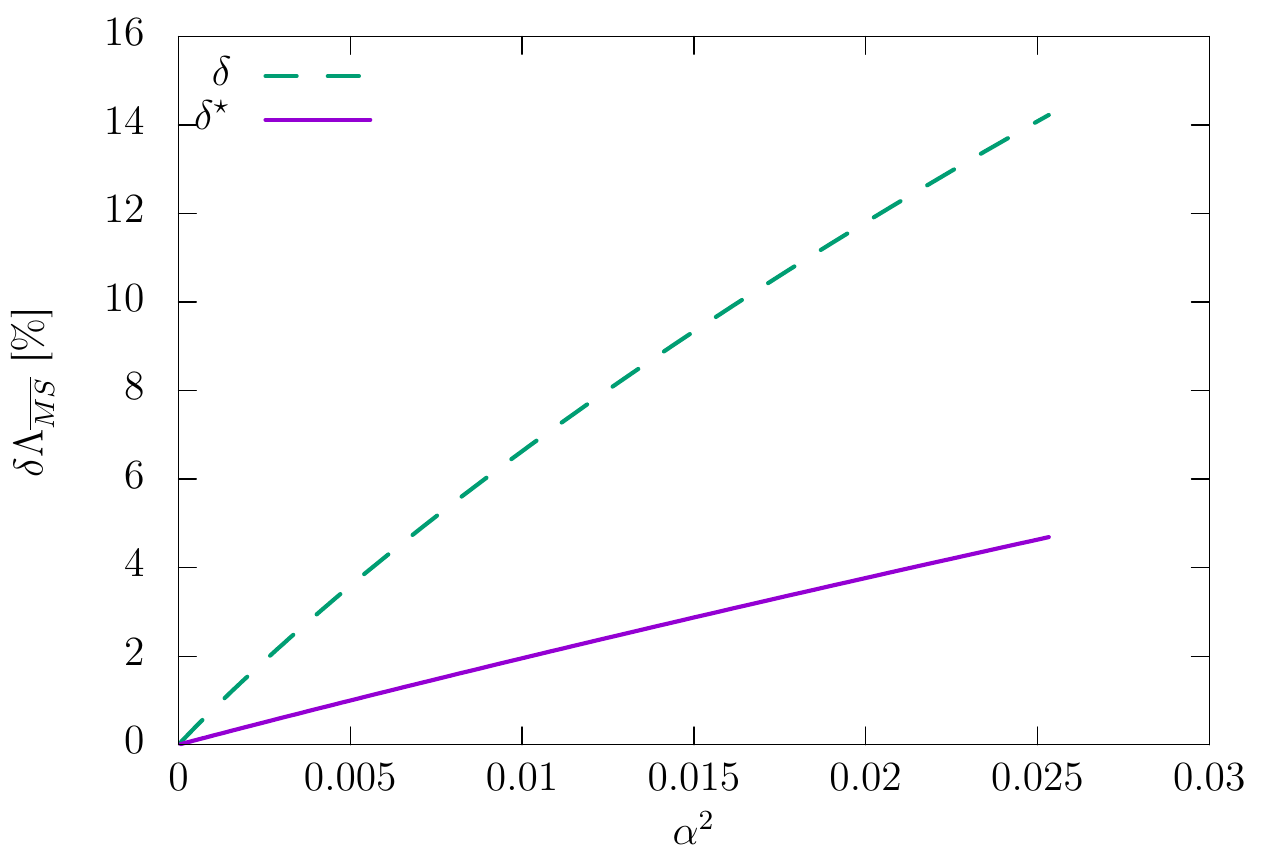}
  \caption{Scale uncertainties for $\Lambda$ associated to different renormalization
  scales for the coupling $\bar g ^2_{\overline{\rm MS} }(s\mu)$.
  The solid curve (labeled $\delta^{\star}$) shows the perturbative
  uncertainty estimated by 
  varying the scale around the scale of fastest apparent convergence. 
  The dashed curve (labeled $\delta$) shows the perturbative
  uncertainty estimated by 
  varying the scale around the physical scale. 
  See text for more details.}
  \label{fig:scale}
\end{wrapfigure}
The approach labeled $\overline{\rm MS} $ in
section~\ref{sec:extraction-lambda} is very close to many
phenomenological extractions of the strong coupling. 
The value of $\bar g ^2_{\overline{\rm MS} }$ is extracted from a
measurement, in this case from the value of the SF coupling
obtained in a simulation, thanks to its perturbative
expansion 
\begin{displaymath}
   \bar g ^2_{\overline{\rm MS} }(s\mu) = \bar g ^2_{\rm SF}(\mu) +
   \frac{c_1(s)}{4\pi} \bar g ^4_{\rm SF}(\mu) + 
   \frac{c_2(s)}{(4\pi)^2} \bar g ^6_{\rm SF}(\mu) + \dots\,.
\end{displaymath}
Different renormalization scales $s\mu$ can be used for each value of
the physical scale $\mu$. 
The differences between different renormalization scales are an
estimate of the truncation errors (i.e. 
an estimate of the $\mathcal O(g_{\rm PT}^{2(l-1)})$ effects in
equation~(\ref{eq:IPT})). 
In particular, in phenomenology, it is very common to vary the
renormalization scale a factor two above/below some chosen value. 

Figure~\ref{fig:scale} shows such estimate of the uncertainties
propagated to the $\Lambda$-parameter. 
$\delta^\star$ is obtained by varying the renormalization scale of a factor two
above/below the scale of fastest apparent convergence (i.e. 
the average difference between the values of $\Lambda$ obtained after using
$s=1$ and $s=2$ and then $s=2$ and $s=4$). 

From figure~\ref{fig:scale} it is clear that that scales uncertainties 
are rather large in the pure gauge theory\footnote{In fact they are even
larger if one varies the scale around the physical scale (by a factor
three), instead of using the value of fastest convergence. 
See figure~\ref{fig:scale}.}.
Even at $\alpha\approx 0.1$, corresponding to the highest scales 
reached in our study, they are around 2\%.
One might question the results of our works, that claim a
significantly smaller uncertainty. 
The key to claim smaller errors than the scale uncertainties lies in
the limit definition of $\Lambda$, equation~(\ref{eq:lamlimit}). 
Once the limit $g_{\rm PT}\to 0$ is properly taken, and its systematic
estimated, one does not need
to talk about the uncertainties at non-zero $g_{\rm PT}$. 
Of course taking such a limit is hard: data at different values
of $g_{\rm PT}$ is required. 
Due to the logarithmic running of the coupling with the physical
energy scale, the apparently innocent limit of
equation~(\ref{eq:lamlimit}) requires to solve a hard multi-scale problem. 
Even with our datasets, that spans a factor 32 in energy scales (a
change in the coupling $g_{\rm PT}^2$ by more than a factor two), we
have seen that some assumptions on the scaling as $g_{\rm PT}\to 0$ are
needed in order to reach the 1.4\% precision on $\Lambda$. 

We consider our approach to treat perturbative uncertainties
very conservative. 
Still, future works in the pure gauge theory might want to explore even
larger energy scales.

%%% Local Variables:
%%% mode: latex
%%% TeX-master: "paper"
%%% End:

\section{Conclusions}
\label{sec:conclusions}

In this work we have examined the main sources of 
uncertainties present in finite-size scaling studies using the Gradient
Flow: the continuum extrapolation and the statistical uncertainties. 
We have argued that scaling violations are a result of exploring
changes in the flow time. This observation has been
supported both by a perturbative study and by non-perturbative
numerical results. 

The determination of the step scaling function $\sigma(u)$, the crucial observable
in step scaling studies, involves both a change in the flow time and a
change in the size of the system.  
We propose to divide the determination of $\sigma(u)$ in two pieces:
first, a change in the renormalization scale at a constant size of the
system (the function $\mathcal J_1$), followed by a change of the size 
of the system at constant renormalization
scale (the function $\mathcal J_2$). 
The advantage is that, according to our hypothesis, only the first
step shows significant cutoff effects. By breaking up the
determination in two pieces, the scaling violations can be studied
much more accurately. Modest datasets allow to explore a change of the
renormalization scale at constant physical volume with lattice
spacings varying by factors 4-6. 
In section~\ref{sec:new-strat-determ} we have seen that this
strategy usually comes at the cost of larger statistical
uncertainties, especially in schemes like the Schr\"odinger Functional 
that break translation invariance and have to deal with
$\mathcal O(a)$ systematic effects.  
Thus, the proposal trades the large systematic associated with the
continuum extrapolations present in many GF studies with a statistical
uncertainty. Since the latter are much easier to control, we think
that the proposed strategy shows a clear advantage. 
In section~\ref{sec:stat-uncert} we have shown that
statistical uncertainties can be well understood and predicted with a
simple model. 

We think that this strategy can shed some light in many problems that
are currently being studied where systematic effects of the continuum
extrapolation are relevant (see for example the recent discussion
in~\cite{Witzel:2019jbe}). A detailed study of the 
scaling violations of $\mathcal J_1$, that according to our hypothesis
are very similar to those of the step scaling function $\sigma$, should
become a standard way to assess the quality of the continuum
determination of the step scaling function. This is specially relevant
for studies of the conformal window, since much less is known about
the logarithmic corrections to scaling in this model, and as we have
shown in section~\ref{sec:non-pert-study} they can have a large effect
in the extrapolations.

We have also re-examined the determination of the
$\Lambda$-parameter in the pure gauge theory. The most crucial step is
the high energy region and the matching with the asymptotic
perturbative regime. We have used the step scaling function with $c=0.2$, determined
using our new proposal. The matching with
perturbation theory is performed in different schemes and using
different procedures. Our datasets allows to match with perturbation
theory at energy scales $\mu_{\rm PT}$ where $\alpha(\mu_{\rm PT})
\lesssim 0.1$. 
Even at this large energy scales the perturbative truncation effects
are large, corresponding to about a $2\%$ uncertainty in $\Lambda$. 
The size of this uncertainty is also confirmed by a scale variation analysis. 
All in all, perturbative errors are large in the pure gauge theory,
making the determination of $\Lambda$ rather challenging in this aspect,
in particular when compared with the corresponding determination in
QCD\footnote{Note however that the coupling runs faster in the pure
  gauge theory. For a fixed range of scales, the pure gauge theory allows to study
larger variations in the coupling.}.
Fortunately, our dataset explores a large range of energy scales, and
gives us the possibility to explore the limit $\alpha(\mu_{\rm PT})\to 0$ 
(corresponding to $\mu_{\rm PT}\to \infty$). 
Once this limit is properly taken, the perturbative uncertainties at
$\alpha(\mu_{\rm PT})$ estimated using scale variation or any other
procedure are irrelevant. Of course taking such a limit is very
challenging. 
The corrections, $\mathcal O(\alpha^2(\mu_{\rm PT}))$, decrease
very slowly due to the logarithmic running of the coupling at high
energies. 

Taking the limit $\alpha_{\rm PT} \to 0$ is very challenging in
a large volume setup due to the finite range of scales that any lattice
simulation can probe. This might explain the difference with some
precise results in large volume~\cite{Kitazawa:2016dsl}, although a
more detailed study is necessary.

Depending on the assumptions made in the extrapolation
$\alpha(\mu_{\rm PT})\to 0$, the uncertainty in $\Lambda$ varies in
the range $1-2\%$. 
All in all, our results show a perfect agreement with the final result
of~\cite{DallaBrida:2019wur} ($\sqrt{8t_0}\times
\Lambda_{\overline{\rm MS}} = 0.6227(98)$), obtained with $c=0.3$,
that quotes an uncertainty $\approx 1.4\%$. We stress once more 
that the method that we used in this work provides a careful
control on the continuum extrapolations, leading to a final uncertainty 
on $\sqrt{8t_0}\times\Lambda_{\overline{\rm MS}}$ of the same size 
of~\cite{DallaBrida:2019wur} even when using a very conservative approach.
Due to the different treatments of perturbation theory and the use of
different schemes, it seems clear that, despite some discrepancies with
other works (see discussion in~\cite{DallaBrida:2019wur}), the
systematic effects in~\cite{DallaBrida:2019wur} are well under
control.
 
%%% Local Variables:
%%% mode: latex
%%% TeX-master: "paper"
%%% End:

\section*{Acknowledgments}
\addcontentsline{toc}{section}{Acknowledgments}

We want to thank specially Rainer Sommer for many useful discussions,
sharing his ideas with us and a careful reading of the manuscript.

We thank M. 
Dalla Brida for his contribution in producing the datasets of
reference~\cite{DallaBrida:2019wur} that was essential for the results
presented here, and E. 
Bribian and M. 
Garcia-Perez for sharing the data of reference~\cite{Bribian:2020xfs}
used in section~\ref{sec:stat-uncert}. 

AR acknowledges financial support from the Generalitat Valenciana
(genT program CIDEGENT/2019/040).

\cleardoublepage

\appendix

\section{Boundary $\mathcal O(a)$ effects}
\label{sec:boundary-mathcal-oa}

The procedure to estimate the $\mathcal O(a)$ cutoff effects is
completely analogous as in~\cite{DallaBrida:2019wur}. 
In fact we use the same datasets. 
We determine numerically the dependence of the GF coupling at
different values of $c=0.2,0.3,0.4$ with the boundary parameter
$c_t$. This is done using simulations at different values of the
improvement parameter $c_t$ close to its 2-loop value
\begin{equation}
  c_t^\star(g_0) = 1 - 0.089g_0^2 - 0.0294 g_0^4 + \mathcal O(g_0^6)\,,\qquad
  (g_0^2 = 6/\beta)\,,
\end{equation}
for lattice sizes $L/a = 8,10,12$. This allows to obtain
\begin{equation}
  \frac{\partial \bar g^2_c}{\partial c_t}\Big|_{c_t = c_t^\star} =
  \frac{a}{L}\left[ a_0(c)\bar g_c^2 + a_1(c) \bar g_c^4 \right]\,.
\end{equation}
Table~\ref{tab:oafit} shows the result of the parameters $a_0(c), a_1(c)$. 
\begin{table}
  \centering
  \begin{tabular}{llll}
    \toprule
    &$c=0.2$&$c=0.3$& $c=0.4$ \\
    \midrule
    $a_0(c)$ &  0.04(3) & -0.14(5) & -0.88(7) \\
    $a_1(c)$ & -0.11(2) & -0.26(3) & -0.43(3) \\
    \bottomrule
  \end{tabular}
  \caption{Values of the fit coefficients $a_0(c)$ and $a_1(c)$ that
    parameterize the sensitivity of $\bar g^2_c$ to the boundary
    improvement coefficient $c_t$.}
  \label{tab:oafit}
\end{table}

Now we need an estimate of how much the true, non-perturbative, value
of $c_t$ differs from its 2-loop value. 
There is no information available, but the extrapolations at constant
$u$ of section~\ref{sec:non-pert-study} completely ignored this linear
effects and the data showed no significant deviation from a $\mathcal
O(a^2)$
scaling. 
This suggests that these effects are smaller than our statistical uncertainties. 
With this insight, a conservative approach consists in using
the full 2-loop contribution as an estimate of the difference between
the 2-loop value and the correct non-perturbative one. 

In summary we add to all our data, in quadratures, the error
\begin{equation}
  \delta \bar g_c^2 = \frac{a}{L}\left[ a_0(c)\bar g_c^2 + a_1(c) \bar g_c^4 \right]\times
  0.0294 g_0^2\,.
\end{equation}
in order to account for a possible mis-tuning of the boundary
$\mathcal O(a)$-improvement parameter $c_t$.

%%% Local Variables:
%%% mode: latex
%%% TeX-master: "paper"
%%% End:

\section{Matching with the SF scheme}
\label{sec:matching-with-sf}

Since the strategy is the same as the one used
in~\cite{DallaBrida:2019wur}, we refer the interested reader to this
reference for more details. 

Here it is enough to say that 
we performed a set of SF simulations with background field and
lattice sizes $L/a = 6, 8, 10, 12, 16$ and at the same values of the
bare coupling $g_0^2=6/\beta$ as the available measurements of the GF
coupling in \emph{twice the lattice size} $L/a=12, 16, 20, 24, 32$. 
We collect between $9\times 10^5$ and $2\times 10^6$ measurements of
the SF coupling $\bar g^2_{\rm SF}$. We further remove all cutoff 
effects up to 2 loops from $\bar g^2_{\rm SF}$ (i.e. 
the leading cutoff effects are $\mathcal O(g_0^6)$).
The non-perturbative data is fitted to a functional form
of the type
\begin{equation}
  \frac{1}{\bar g^2_{\rm SF}(\mu)} -   \frac{1}{\bar g^2_{c=0.3}(\mu/(2c))} =
  f(u) + \left( \frac{a}{L} \right)^2 \tilde \rho(u), \qquad
  (u = \bar g^2_{c=0.3}(\mu/(2c)))\,,
\end{equation}
where both $f(u)$ and $\tilde \rho(u)$ are simple polynomials in $u$. 
For the GF coupling we use the measurements at $c=0.3$. 
The matching with other values of $c$ is performed thanks to the
knowledge of the function $\mathcal J_3$ of equation~(\ref{eq:J3def}). 

To quote all our results we use the same fit used in
reference~\cite{DallaBrida:2019wur}, where the coarser lattice is
dropped from the fit and the functions $f(u),\tilde \rho(u)$ are
degree two polynomials.

%%% Local Variables:
%%% mode: latex
%%% TeX-master: "paper"
%%% End:

\clearpage
%\section*{References}
\addcontentsline{toc}{section}{References}
\bibliography{/home/alberto/docs/bib/math,/home/alberto/docs/bib/campos,/home/alberto/docs/bib/fisica,/home/alberto/docs/bib/computing}

\end{document}